\documentclass[iop,apj]{emulateapj}

\usepackage{apjfonts}
\usepackage{natbib}

\newcommand{\vc}[1]{\textbf{\em #1}}

\shorttitle{MHD Reconnection and Particle Acceleration}
\shortauthors{Kowal, de Gouveia Dal Pino \& Lazarian}

\begin{document}

\title{Magnetohydrodynamic Simulations of Reconnection and Particle
       Acceleration: Three-Dimensional Effects}
\author{Grzegorz Kowal$^{1,3}$, E. M. de Gouveia Dal Pino$^{1}$
     \& A. Lazarian$^{2}$}
\affil{$^{1}$Instituto de Astronomia, Geof\'\i sica e Ci\^encias Atmosf\'ericas,
             Universidade de S\~ao Paulo, Rua do Mat\~ao 1226, 05508-900
           , S\~ao Paulo, Brazil}
\affil{$^{2}$Department of Astronomy, University of Wisconsin,
             475 North Charter Street, Madison, WI 53706, USA}
\affil{$^{3}$Obserwatorium Astronomiczne, Uniwersystet Jagiello\'nski,
             ul. Orla 171, 30-244 Krak\'ow, Poland}
\email{kowal@astro.iag.usp.br; dalpino@astro.iag.usp.br;
       alazarian@facstaff.wisc.edu}

\begin{abstract}
The magnetic fields can change their topology through a process known as
magnetic reconnection.  This process in not only important for understanding the
origin and evolution of the large-scale magnetic field, but is seen as a
possibly efficient particle accelerator producing cosmic rays mainly through the
first order Fermi process.  In this work we study the properties of particle
acceleration in reconnection zones and show that the velocity component parallel
to the magnetic field of test particles inserted in nearly non-resistive
magnetohydrodynamic (MHD) domains of reconnection without including kinetic
effects, such as pressure anisotropy, the Hall term, or anomalous effects,
increases exponentially.  Also, the acceleration of the perpendicular component
is always possible in such models.  We have found that within contracting
magnetic islands or current sheets the particles accelerate predominantly
through the first order Fermi process, as previously described, while outside
the current sheets and islands the particles experience mostly drift
acceleration due to magnetic fields gradients.  Considering two dimensional MHD
models without a guide field, we find that the parallel acceleration stops at
some level.  This saturation effect is however removed in the presence of an
out-of-plane guide field or in three dimensional models.  Therefore, we stress
the importance of the guide field and fully three dimensional studies for a
complete understanding of the process of particle acceleration in astrophysical
reconnection environments.
\end{abstract}

\section{Introduction}

Acceleration of energetic particles is important for a wide range of
astrophysical environments, from stellar magnetospheres, accretion disk/jet
systems, supernova remnants and gamma ray bursts to clusters of galaxies.
Several mechanisms for particle acceleration have been discussed in the
literature which include varying magnetic fields in compact sources, stochastic
second order Fermi process in turbulent interstellar and intracluster media, and
the first order Fermi process behind shocks.  An alternative, less explored
mechanism so far, involves particle acceleration in magnetic reconnection sites,
and this will be the focus of the present work which is a first of a series of
papers on this subject.  For a comprehensive recent review on particle
acceleration mechanisms the reader is referred to \cite{melrose09}.

Magnetic reconnection may occur when two magnetic fluxes of opposite polarity
encounter each other.  In the presence of finite magnetic resistivity, the
converging magnetic lines annihilate at the discontinuity surface and a current
sheet forms there.  It is common knowledge that magnetic fields stay frozen in
highly conductive fluids.  Estimates show that Ohmic diffusion for astrophysical
scales is absolutely negligible, contributing to the strongly entrenched view
that interacting magnetic fields of opposite polarity change their topology at
low speeds.  This is determined by the magnetic reconnection speed, and the
famous example of two oppositely directed magnetic fluxes in Ohmic contact
undergoing slow Sweet-Parker \citep{sweet58,parker57} process of magnetic flux
annihilation is frequently invoked.  In a more generic context, the magnetic
reconnection rate is the speed at which two magnetic flux tubes, which are
pushed against each other, can pass through altering the initial field topology.
 This is a situation that can be frequently encountered in astrophysical fluids
with complex motions.

In the Sweet-Parker model, it has been shown that particles can accelerate due
to the induced electric field in the reconnection zone \citep{litvinenko03}.This
$one-shot$ acceleration process, however, is constrained by the narrow thickness
of the acceleration zone which has to be larger than the particle Larmor radius
and by the strength of the magnetic field.  Therefore, the efficiency of this
process is rather limited.  Besides, it also does not predict a power-law
spectrum, as generally observed for cosmic rays.

Observations have always been suggestive that magnetic reconnection can happen
at a high speed in some circumstances, in spite of the theoretical difficulties
in explaining it.  For instance, the phenomenon of solar flares suggests that
magnetic reconnection should be first slow in order to ensure the accumulation
of magnetic flux and then suddenly become fast in order to explain the observed
fast release of energy.  A model that can naturally explain this and other
observational manifestations of magnetic reconnection was proposed by
\cite{lazarian99}.  The model appeals to the ubiquitous astrophysical turbulence
as a universal trigger and controller of fast reconnection.  The predictions of
the model have been successfully tested in numerical simulations by
\cite{kowal09} which confirmed that this speed is of the order of the Alfv\'en
speed in the presence of weakly stochastic magnetic field fluctuations.

An important consequence of fast reconnection of turbulent magnetic
fields\footnote{This model of fast reconnection does not impose limits on the
amplitude of magnetic perturbations.  They can be small and the magnetic flux
tubes require just to have a weak turbulent noise in order to produce fast
magnetic reconnection, i.e., the reconnection which does not depend on Ohmic
resistivity.} is the formation of a thick volume filled with reconnected small
magnetic flux loops.  Now, if such turbulent flow is immersed in a current sheet
formed by two large scale converging magnetic flux tubes (such as, e.g., in the
Sweet-Parker configuration), then the three-dimensional magnetic fluctuations or
loops will contract and scatter test particles, presenting favorable conditions
for energetic particle acceleration in a first order Fermi process.  In other
words, the particle may bounce back and forth between these converging magnetic
mirrors formed by oppositely directed magnetic fluxes moving towards each other
with the velocity corresponding to the reconnection speed.  This has been first
described in \cite{degouveia05} \cite[see also][]{lazarian05} for the situation
when there is no back reaction of the accelerated particles on the reconnected
magnetic flux\footnote{This mechanism is in contrast to the second order Fermi
acceleration which is frequently discussed in terms of the particle acceleration
by turbulence generated by reconnection \citep{larosa06}, \cite[see also][in
prep.]{kowal11}.}.  Later, \cite{drake06} appealed to a similar process, but
within a collisionless reconnection scenario.  In their model, the contraction
of two-dimensional loops is controlled by the firehose instability that arises
in the particle-in-cell (PIC) simulations containing both electrons and ions
\citep{drake10}.  In the present work, we show that this type of acceleration
can be present in a pure magnetohydrodynamical scenario as well, where the
pressure is fully isotropic and the contraction of islands is determined by
their interactions.

Magnetic reconnection is ubiquitous in astrophysical circumstances and therefore
it is expected to induce acceleration of particles in a wide range of
astrophysical environments including galactic and extragalactic ones.  For
instance, the process has been already discussed for the production of
ultra-high energy cosmic rays \citep{degouveia00,degouveia01}, acceleration of
particles in gamma ray bursts \citep{lazarian03,zhang11}, microquasars
\citep{degouveia05} and astrophysical jet-accretion disks in general
\citep{degouveia10a,degouveia10b}.  In particular, in the case of relativistic
jets, a diagram of the magnetic energy rate released by violent reconnection as
a function of the black hole (BH) mass spanning $10^9$ orders of magnitude was
derived and demonstrates that the magnetic reconnection power is more than
sufficient to explain the observed radio outbursts, from microquasars to low
luminous active galactic nuclei (AGN) \citep{degouveia10a}.

More recently, the acceleration in reconnection regions has obtained
observational support.  It was suggested in \cite{lazarian09} that anomalous
cosmic rays measured by Voyagers are, in fact, accelerated in the reconnection
regions of the magnetopause \citep[see also][]{drake10}.  Such a model explains
why Voyagers did not see any signatures of acceleration passing the Solar system
termination shock.  In a separate development, \cite{lazarian10} have appealed
to the energetic particle acceleration in the wake produced as the Solar system
moves through interstellar gas to explain the excess of cosmic rays of the range
of both sub-TeV and multi-TeV energies in the direction of the magnetotail.
Note, that due to the 11-year solar magnetic cycle the accumulation of magnetic
reversals in the wake is unavoidable.

The implications of the acceleration process in reconnection sites are expected
to be even much wider.  Numerical two-dimensional (2D) simulations recently
presented in \cite{drake10} confirmed the high efficiency of particle
acceleration in regions of magnetic reconnection.  However, we will show here
that the process of acceleration happens rather differently in two and three
dimensional (3D) situations.  The 3D geometry shows a wider variety of
acceleration regimes and this calls for much more detailed studies of the
acceleration.

In Section~\ref{sec:methods} we describe the methodology of the particle
acceleration studies presented here.  In Section~\ref{sec:results} we show the
results obtained from these studies, in Sections~\ref{sec:discussion} and
\ref{sec:summary} we discuss the results and draw the main conclusions of this
work.

\section{Methodology}
\label{sec:methods}

In order to study particle acceleration in magnetic reconnection sites, we
performed numerical simulations solving the isothermal magnetohydrodynamic (MHD)
equations in two and three dimensions (2D and 3D, respectively).  To compare our
MHD results with those obtained with the Particle in Cell (PIC) code in
\cite{drake10} we intentionally, in our 2D simulations, have reproduced their
set up of eight Harris current sheets in a periodic box.  The initial density
profile is chosen in such a way that the total (gas plus magnetic) pressure is
uniform. Initially, we imposed random weak velocity fluctuations to this
environment in order to enable spontaneous reconnection events and the
development of magnetic islands. We evolved the system in 2D (both without and
with a guide field along the third dimension defined by Z direction) and in 3D
domains.  After the development of the magnetic loops (which in a 2D geometry
form islands) in an underlying plasma configuration with defined density,
velocity and magnetic field profiles, we injected test particles at a given
snapshot and integrated their trajectories solving the equation of motion for
each charged particle
\begin{equation}
 \frac{d}{d t} \left( \gamma m \vc{u} \right) = q \left( \vc{E} + \vc{u} \times
\vc{B} \right) , \label{eq:ptrajectory}
\end{equation}
where $m$, $q$ and $\vc{u}$ are the particle mass, electric charge and velocity,
respectively, $\vc{E}$ and $\vc{B}$ are the electric and magnetic fields,
respectively, $\gamma \equiv \left( 1 - u^2 / c^2 \right)^{-1/2}$ is the Lorentz
factor, and $c$ is the speed of light.  In the studies below we assume that the
charged particles are protons.

\begin{figure*}[ht]
 \center
 \includegraphics[width=\textwidth]{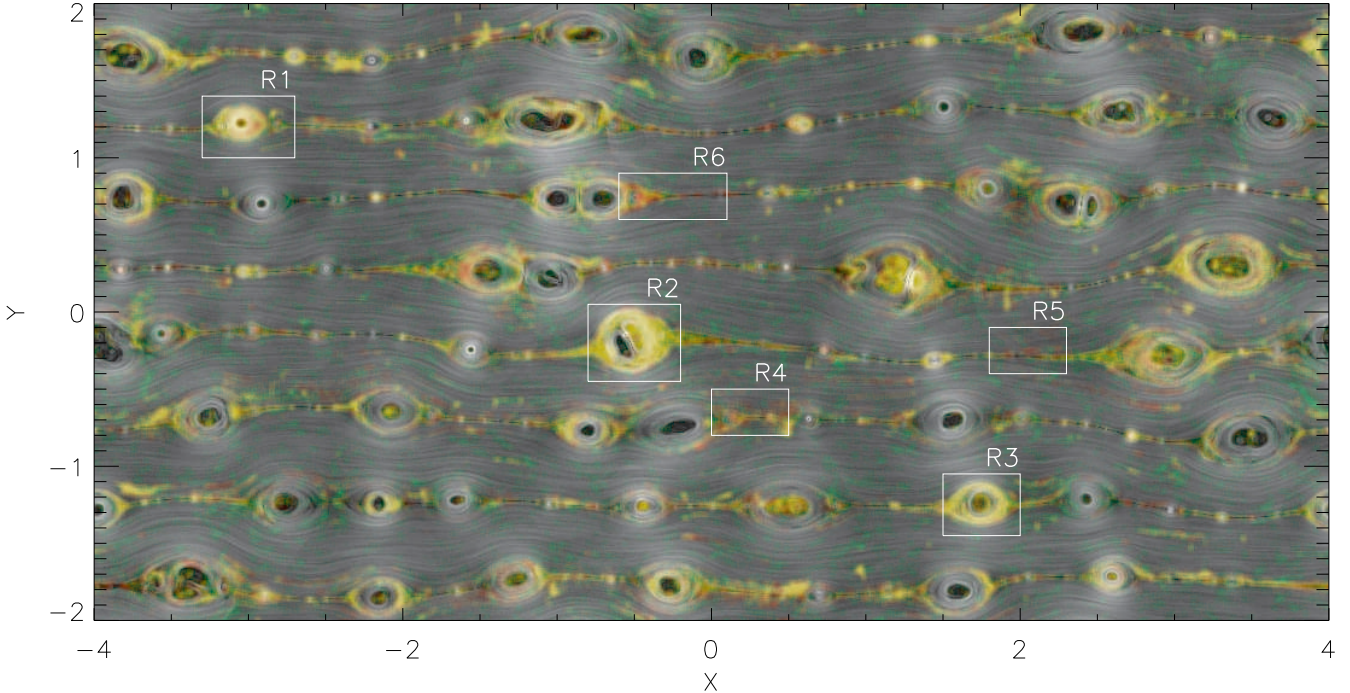}
 \caption{Topology of the magnetic field represented as the gray texture with
semi-transparent color maps representing locations where the parallel and
perpendicular particle velocity components are accelerated for a 2D model with
$B_z = 0.0$ at time $6.0$ in the code units.  The red and green colors
correspond to regions where either parallel or perpendicular acceleration
occurs, respectively, while the yellow color shows locations where both types of
acceleration occur.  The parallel component increases in the contracting islands
and in the current sheets as well, while the perpendicular component increases
mostly in the regions between current sheets.  White boxes show regions that are
more carefully analyzed in this paper.  The simulation was performed with the
resolution 8192x4096.  We injected 10,000 test particles in this snapshot with
the initial thermal distribution with a temperature corresponding to the sound
speed of the MHD model.  \label{fig:locations}}
\end{figure*}

In the MHD simulations the electric field $\vc{E}$ is generated either by the
magnetized plasma or by resistive effects and can be obtained directly from the
Ohm's law equation, i.e.,
\begin{equation}
 \vc{E} = - \vc{v} \times \vc{B} + \eta \vc{j} , \label{eq:efield}
\end{equation}
where $\vc{v}$ is the plasma velocity, $\vc{j} \equiv \nabla \times \vc{B}$ is
the current density, and $\eta$ is the Ohmic resistivity coefficient.

In our studies we are not interested in the acceleration by the electric field
resulting from resistivity effects, therefore we have neglected the last term in
Equation~(\ref{eq:efield}) in the trajectory integration.  The charged particles
then feel an electric field $\vc{v} \times \vc{B}$ besides the magnetic field.
Substituting the Ohm's law, the equation of motion can be rewritten as
\begin{equation}
 \frac{d}{d t} \left( \gamma m \vc{u} \right) = q \left[ \left( \vc{u} - \vc{v}
\right) \times \vc{B} \right] . \label{eq:trajectory}
\end{equation}

The particle equation of motion (Eq.~\ref{eq:trajectory}) was integrated using
the 6$^{th}$ order implicit Runge-Kutta-Gauss (RKG) method
\cite[see][e.g.]{sanz-serna94} with a fixed time step $dt=10^{-7}$.  The RKG
methods are known to conserve the particle energy and momentum in very long
integrations \citep{sanz-serna94} in contrast to the standard 4$^{th}$ order
Runge-Kutta method with the adaptive time step based on the 5$^{th}$ order error
estimator \cite[see][e.g.]{press92}, which is very commonly used.

The interpolation of the local values of the plasma velocity $\vc{v}$ and
magnetic field $\vc{B}$ at each step of the integration has been done using
cubic interpolation \citep{lekien05} with our own discontinuity detector based
on a total variation diminishing (TVD) limiter.  We performed also the
integration using linear interpolation which gave us essentially the same
statistical results, but with much faster integration times.

In the current study we do not include particle energy losses, therefore test
particles can gain or lose energy only through the interactions with the moving
magnetized plasma and its fluctuations.  The inclusion of radiative and
non-radiative losses and also the back reaction of the particles on the plasma
is planned for future studies.

For simplicity, we assume the speed of light to be 20 times the Alfv\'en speed
$V_A$, which defines our plasma in a non-relativistic regime.  The mean density
is assumed to be 1 atomic mass unit per cubic centimeter which is compatible
with the diffuse interstellar medium (ISM) density.  The results are presented
in units normalized by the assumed light speed and time unit (which is 1 hour in
our simulations).

\section{Results}
\label{sec:results}

\subsection{Parallel and Perpendicular Acceleration in 2D Models}
\label{sec:multi-layer}

Figure~\ref{fig:locations} presents an evolved 2D configuration of the magnetic
field structure with magnetic islands.  We clearly see the merging of islands in
some locations and the resulting deformations and/or contractions which provide
appropriate conditions for particle acceleration in a similar way to the results
obtained with the PIC code in \cite[][and references therein]{drake10}.  In
general, we distinguish two kinds of accelerating zones where one or both
velocity components (parallel and perpendicular to the magnetic field) can
increase.

In Figure~\ref{fig:locations}, in order to demonstrate the zones which are
favorable for a particular type of acceleration, we superimposed on top of the
gray texture which represents the magnetic topology, a semitransparent color map
showing the locations where acceleration increases the parallel (red) or the
perpendicular (green) velocity component.  In yellow areas, both types of
acceleration can occur depending on the speed and direction of the test particle
with respect to the orientation of the magnetic field lines.

We see that the increase of the parallel velocity component is mostly observed
within deformed islands and in current sheets (see the red and yellow zones in
Figure~\ref{fig:locations}), while the increase of the perpendicular component
is observed mostly near and within the islands and between current sheets (see
the green and yellow zones in Figure~\ref{fig:locations}).  This complex
behavior is related to the degree of island deformation and the particle
direction and speed.  If the island contracts, the particle can increase its
parallel component, however, when the island increases its deformation the
preferable acceleration usually takes place in the perpendicular direction.
Interestingly, we can find islands in which the acceleration does not happen at
all (see, e.g., the islands above and below the region R3 in
Figure~\ref{fig:locations}).  This is due to the fact that these islands are not
undergoing contraction or deformation processes.

\subsection{Distribution of the Accelerated Particles}
\label{sec:acc_dist}

\begin{figure}[t]
 \center
 \includegraphics[width=0.5\textwidth]{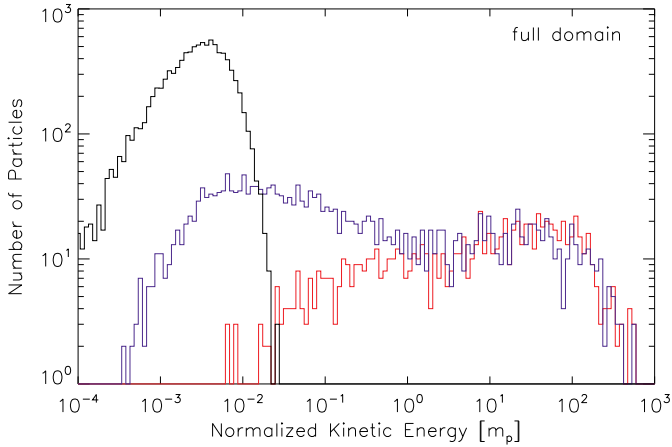}
 \caption{Particle energy distributions in the full domain of
Figure~\ref{fig:locations}.  Blue and red histograms show the number of
particles accelerating their perpendicular and parallel velocity components,
respectively, at one hour after the particles injection in the system.  The
black line exhibits the initial thermal distribution of the injected particles.
\label{fig:particle_histogram}}
\end{figure}

In Figure~\ref{fig:particle_histogram} we present the energy distribution of all
particles of the domain of Figure~\ref{fig:locations} at a time corresponding to
1 hour after the injection, when the particles are still accelerating.  The red
and blue lines show the number of particles which increase their parallel and
perpendicular velocity components, respectively, as a function of the kinetic
energy.  In addition, we show the initial thermal distribution of particles
(black line).

In the particle energy distribution corresponding to the perpendicular
acceleration component (blue), we distinguish two modes, one in the low energy
range which is related to the initially injected thermal distribution with an
extended non-thermal tail, and the other corresponding to the high energy range.
 For the particle energy distribution corresponding to the parallel acceleration
component, we see only one distribution mode with a maximum at high energies. In
the low energy range with energies smaller than the particle rest mass energy
(around $10^{-2}$), there is a clear dominance of the perpendicular
acceleration.  This creates an anisotropy between the two velocity distributions
with respect to the magnetic field.  However, once the particles become
relativistic and their kinetic energies become comparable to or larger than the
particle rest mass energy, the anisotropy disappears and the acceleration in
both directions is equally efficient.

\begin{figure*}[ht]
 \center
 \includegraphics[width=0.32\textwidth]{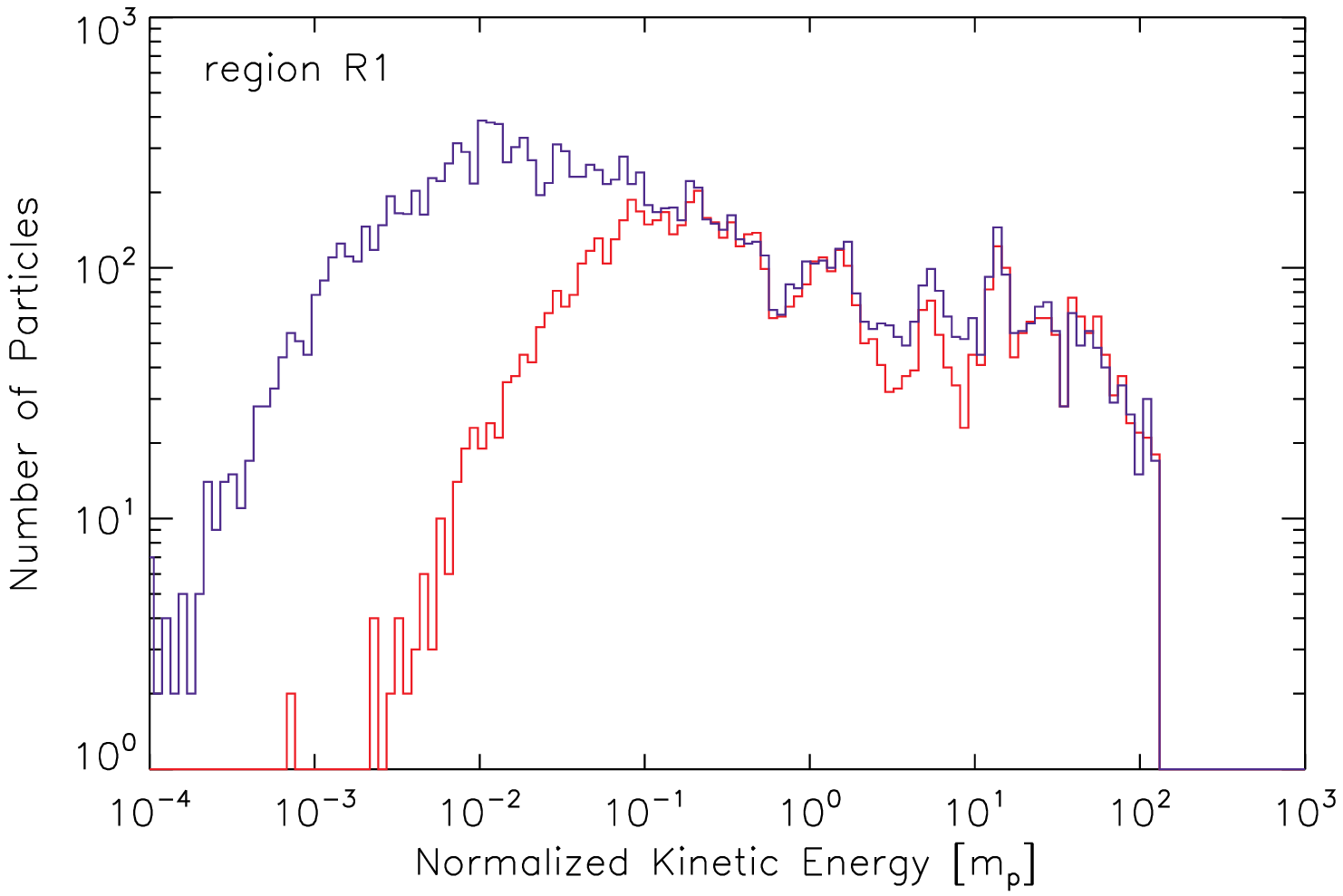}
 \includegraphics[width=0.32\textwidth]{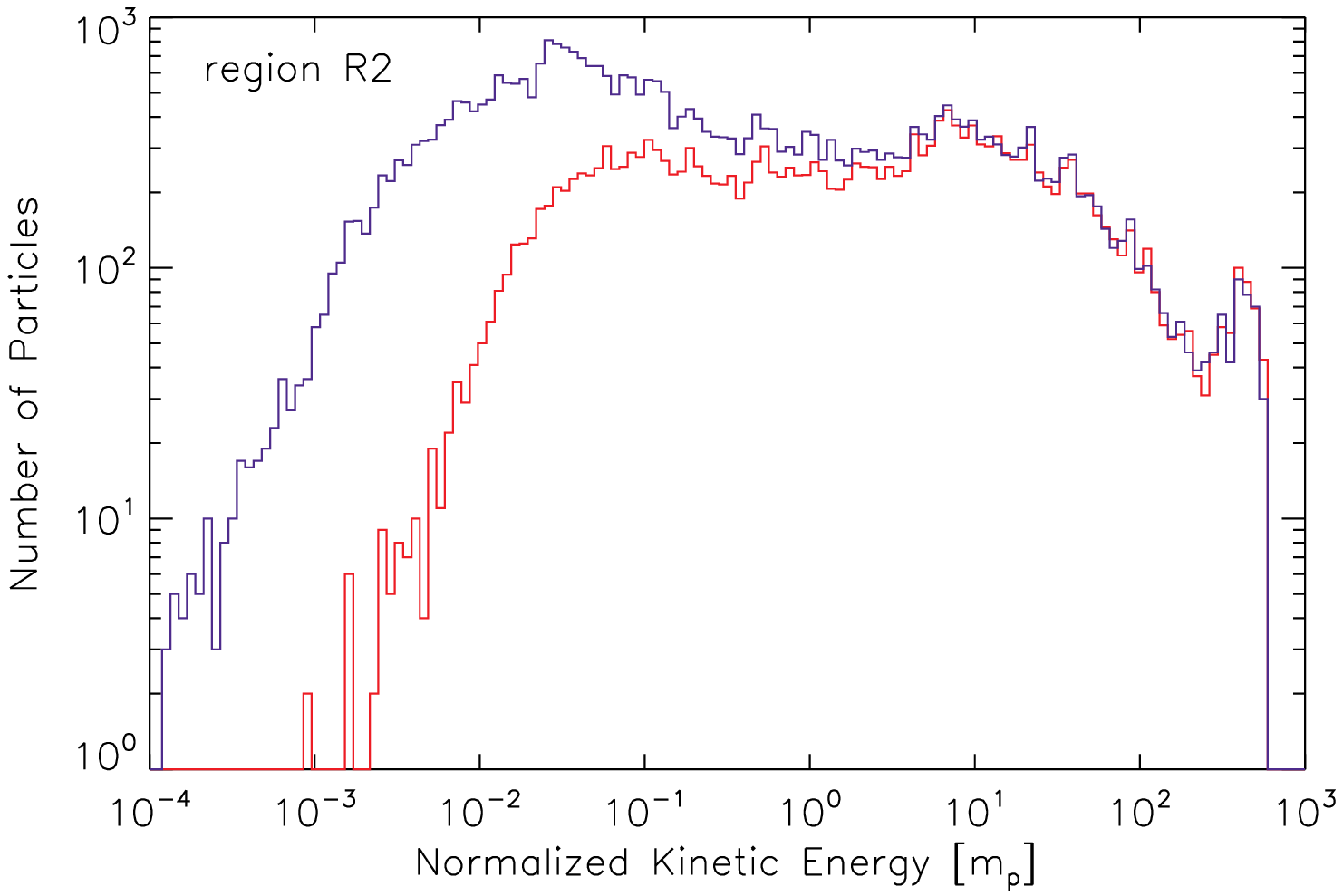}
 \includegraphics[width=0.32\textwidth]{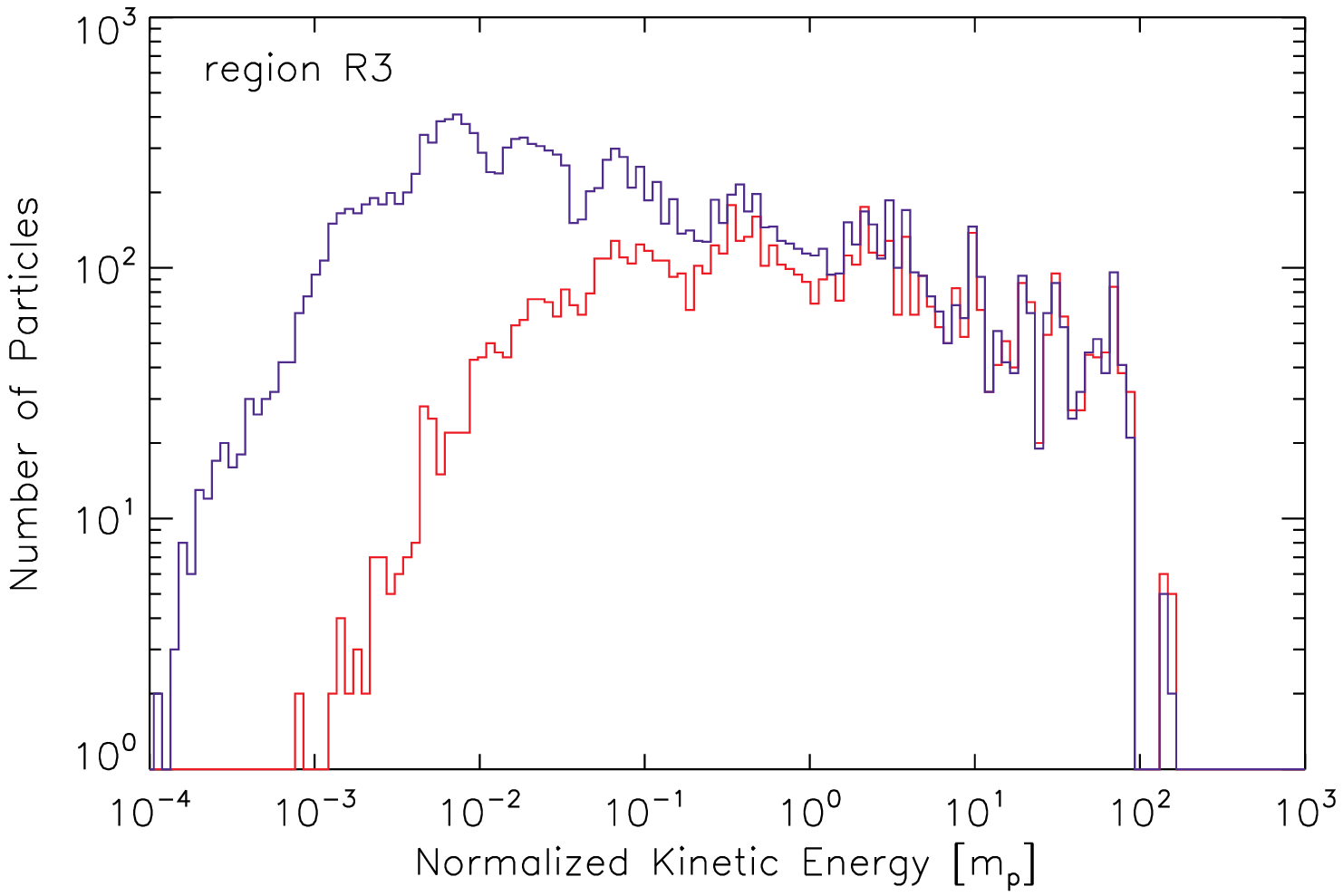}
 \includegraphics[width=0.32\textwidth]{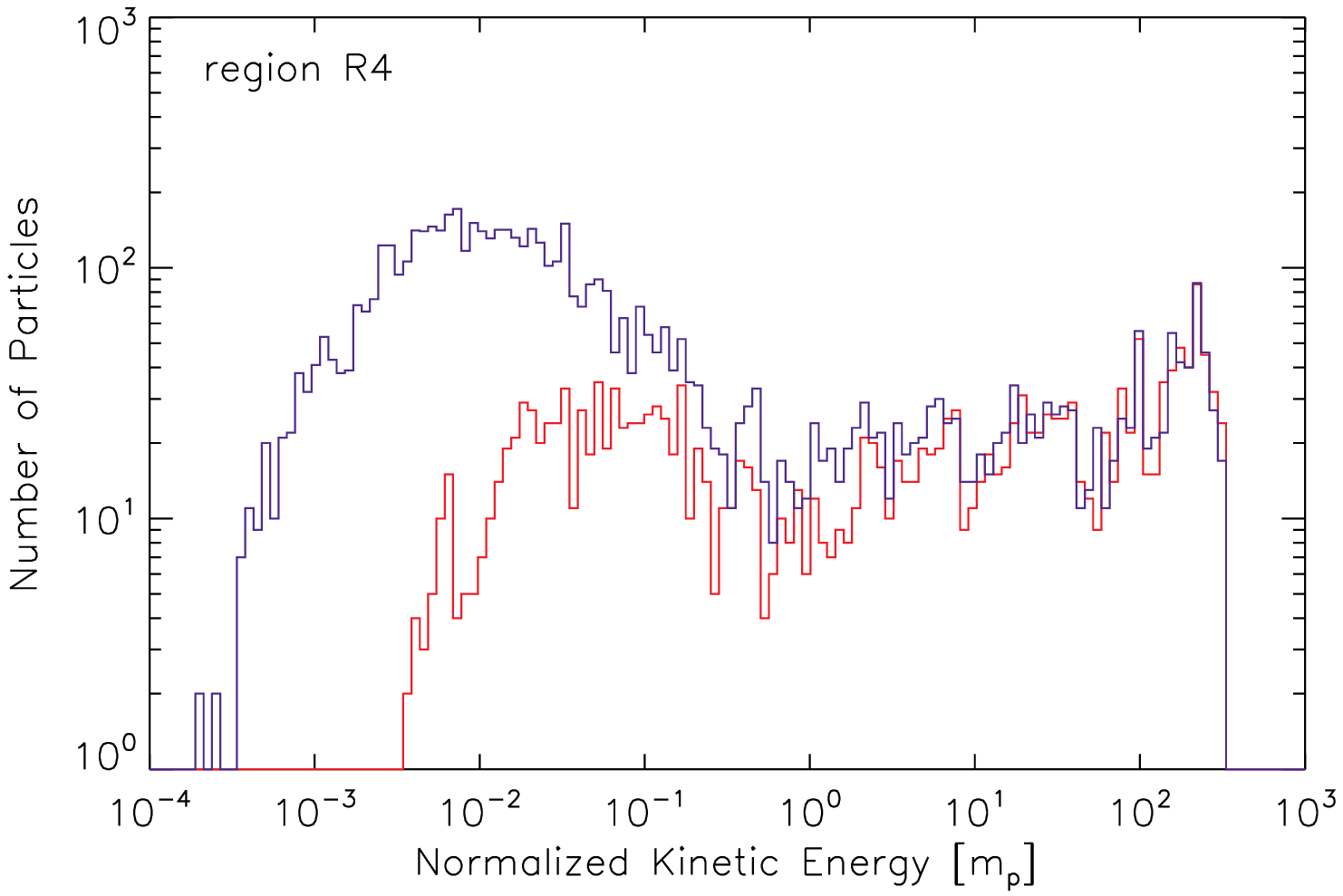}
 \includegraphics[width=0.32\textwidth]{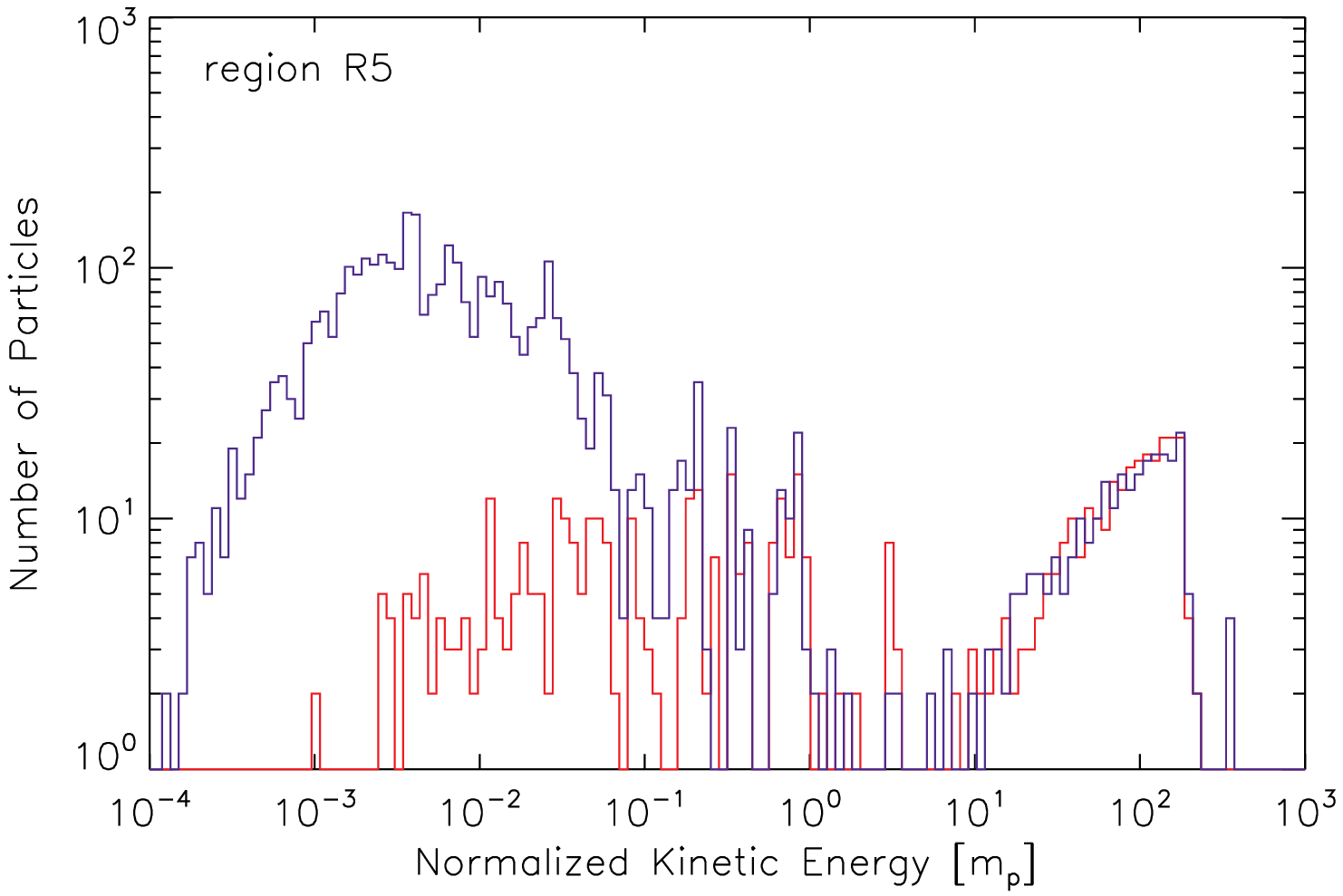}
 \includegraphics[width=0.32\textwidth]{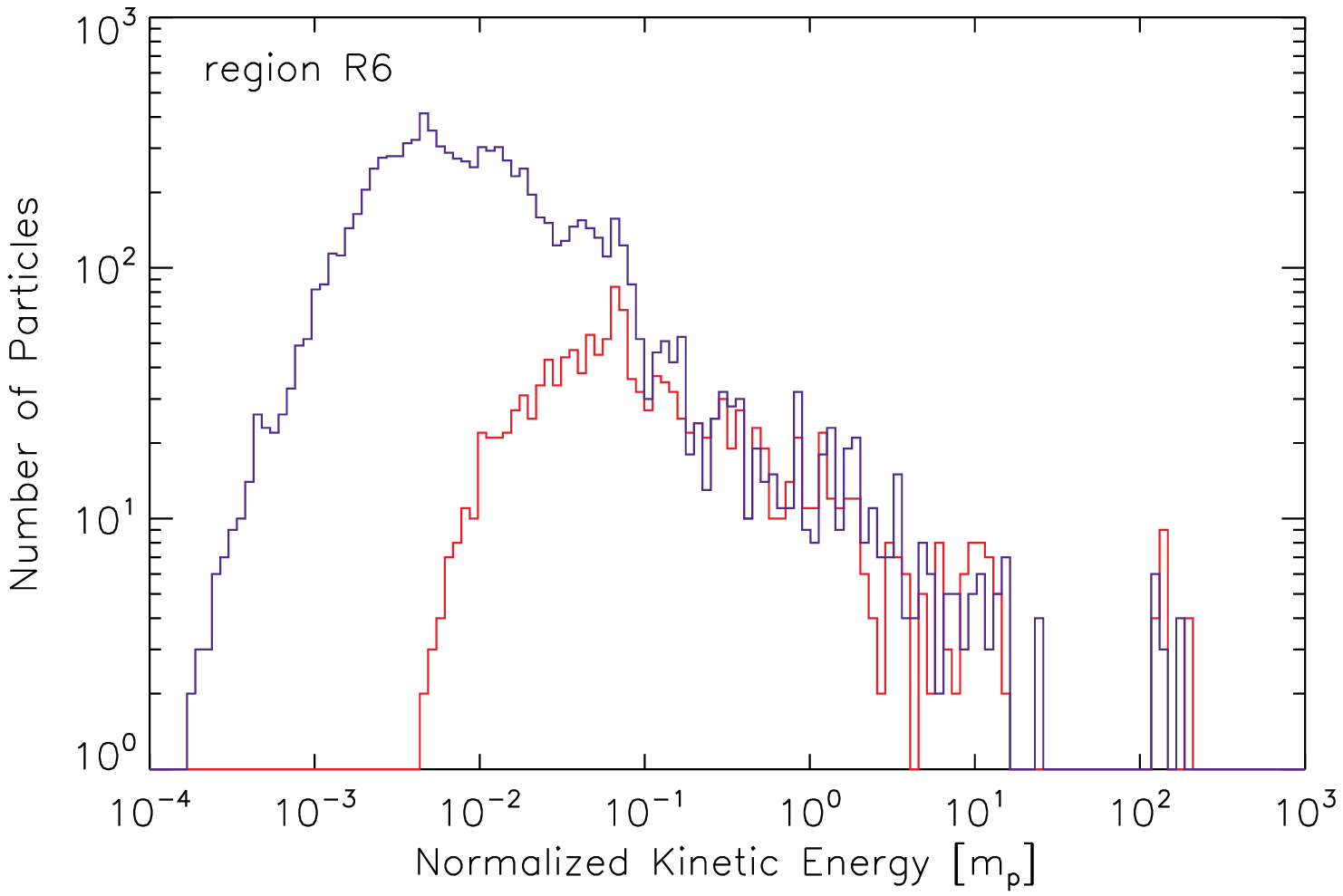}
 \caption{Particle energy distributions in the selected regions of
Figure~\ref{fig:locations}.  Blue and red histograms correspond to the number of
particles which accelerated their perpendicular and parallel velocity
components, respectively. These histograms are created for all accelerating
particles passing through this region from the injection moment until 1 hour.
\label{fig:particle_histograms}}
\end{figure*}

Figure~\ref{fig:particle_histograms} presents the energy distributions of
accelerated particles in the selected regions R1 to R6, as shown in
Figure~\ref{fig:locations}.  Regions R1 to R3 correspond to islands where both
parallel and perpendicular accelerations are observed, while regions R4 to R6
correspond to current sheets.

The energy distributions for regions R1 to R3 are shown in the top panels of
Figure~\ref{fig:particle_histograms}.  We notice a dominance of the
perpendicular acceleration in the low energy range, similarly as seen in
Figure~\ref{fig:particle_histogram}.  As we move to higher energies we see a
trend to form a saturated tail.  The efficiency of the acceleration of the
parallel component increases with the particle energy and at the highest
energies the number of particles accelerating in the parallel and perpendicular
directions is comparable.

In the bottom panels of Figure~\ref{fig:particle_histograms} we present the
particle energy distributions for regions within current sheets which are
labeled as R4, R5, and R6 in Figure~\ref{fig:locations}.  In these cases, we
also observe an anisotropy between the perpendicular and parallel components in
the low energy range with a dominance of the perpendicular acceleration.  As in
the case of the whole domain, this anisotropy decreases with increasing energy
and, beyond values around the particle rest mass energy both component
distributions become comparable.  The high energy range tails of these
distributions are steeper than those of the top diagrams of the figure which
correspond to magnetic islands. This is especially well seen in the energy
distribution of region R6.  This indicates that the number of accumulated
particles in the high energy range is smaller and therefore the acceleration,
although still significant, is less efficient in current sheets than within the
islands.  This is apparently due to the fact that (at least in the 2D case) the
islands are able to retain the particles trapped for longer times.  What could
be the dominant mechanism(s) for the acceleration in these zones?

The acceleration in the contracting/deforming islands is suggestive of first
order Fermi processes with the particles bouncing back and forth between
converging mirrors as described in \cite{degouveia05} and \cite{drake10}, while
within and between the current sheets the acceleration mechanism is not as
clear.  In the following sections, we will examine the acceleration mechanism in
these different sites in more detail.

\subsection{Acceleration in Contracting Islands}
\label{sec:island}

\begin{figure*}[ht]
 \center
 \includegraphics[width=0.40\textwidth]{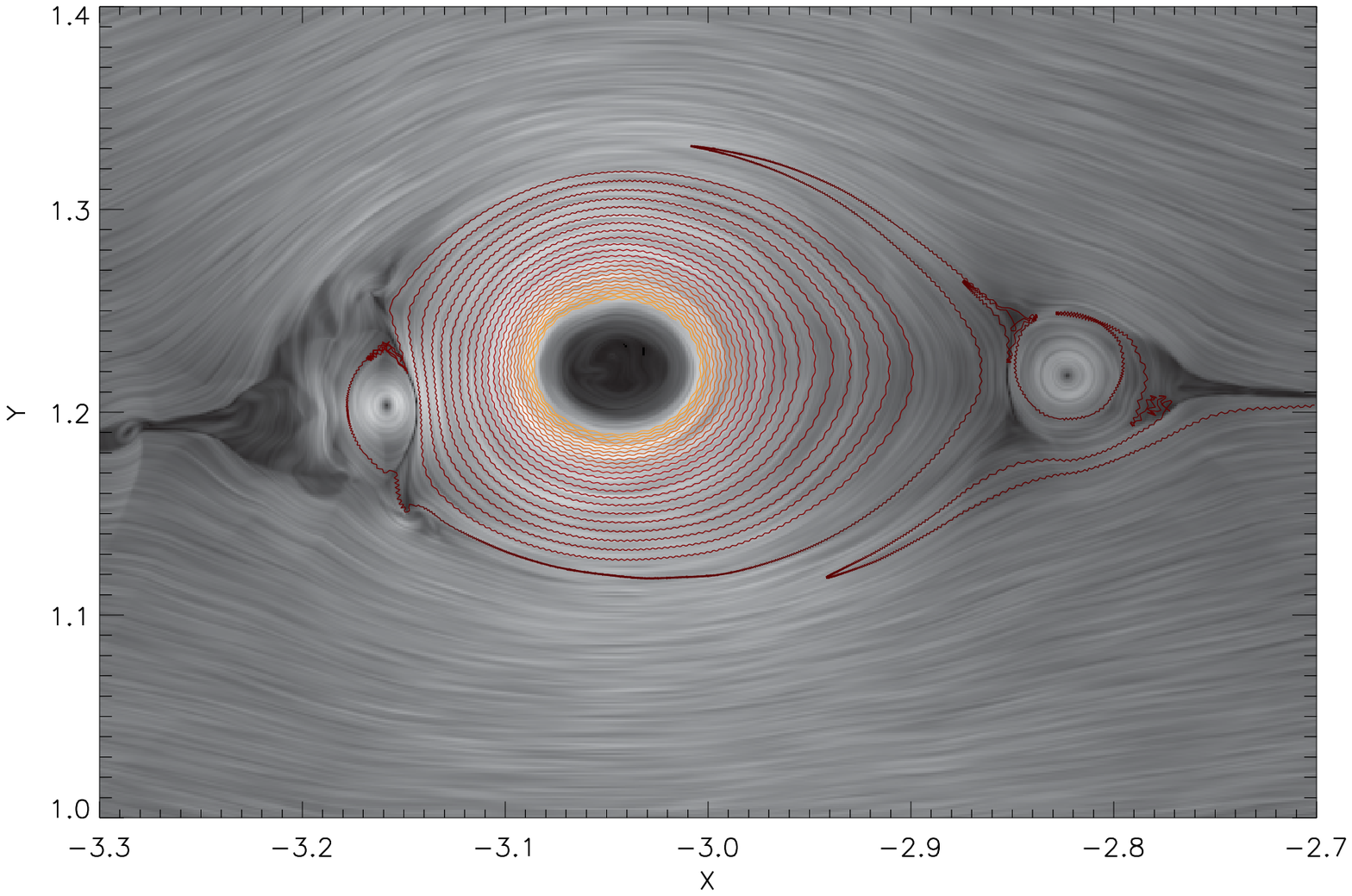}
 \includegraphics[width=0.30\textwidth]{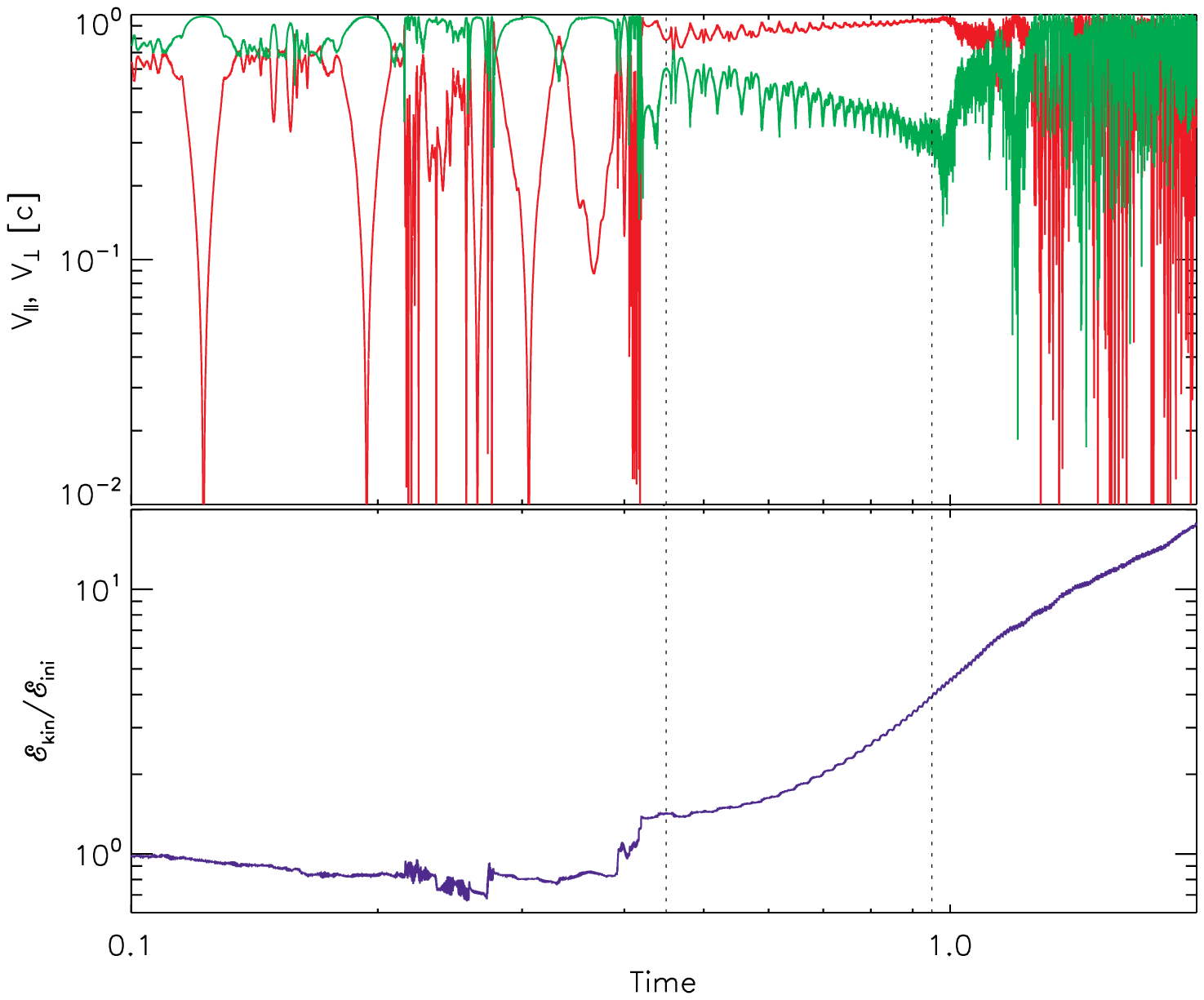}
 \includegraphics[width=0.25\textwidth]{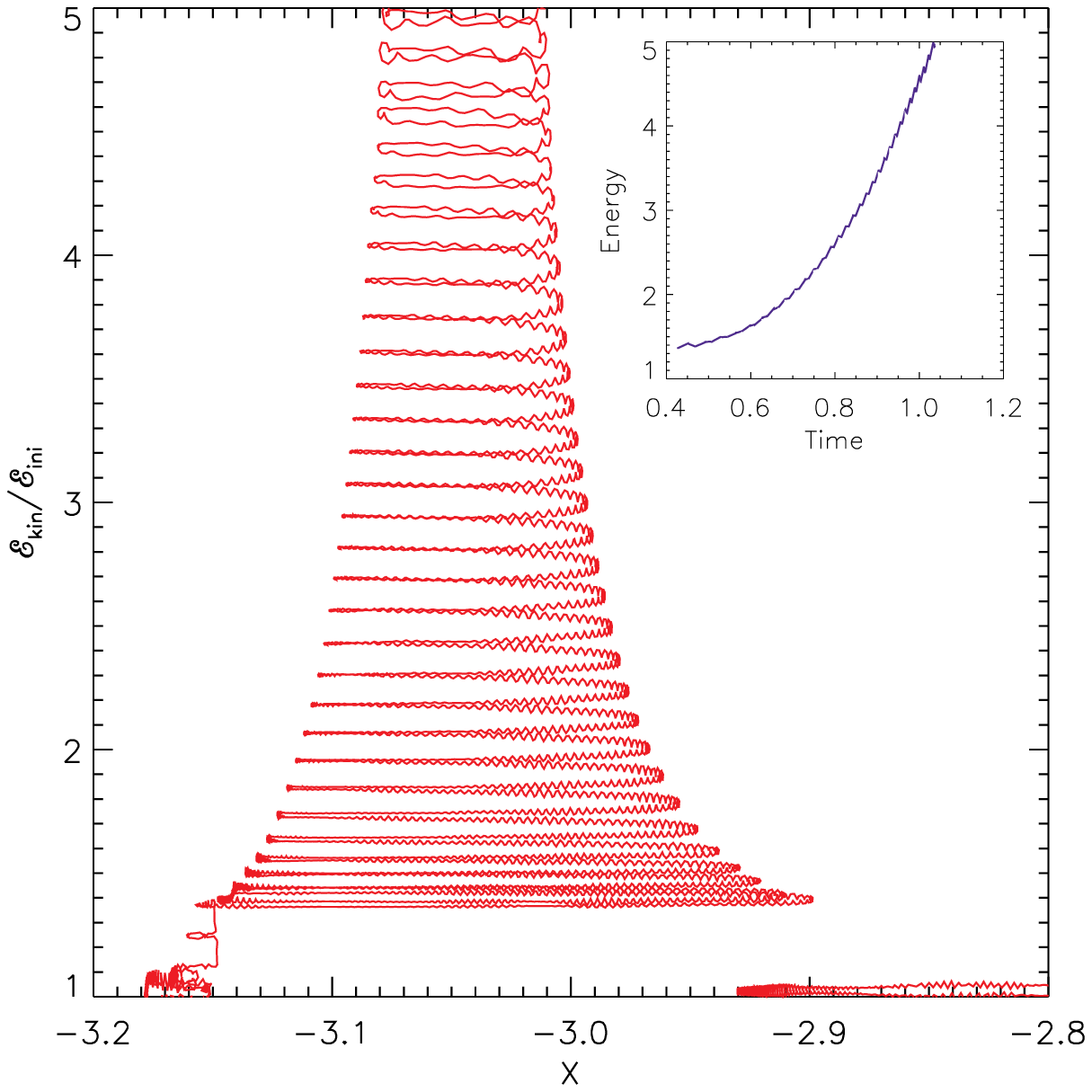}
 \caption{The case of a contracting island where the particles accelerate
efficiently (region R1 of Fig.~\ref{fig:locations}).  In the left panel we show
the trajectory of a test proton trapped in a contracting island.  We see two
small magnetic islands on both sides of the central elongated island which are
merging with it.  This process results in the contraction of the central island.
In the middle panel we show the evolution of the parallel and perpendicular
speeds of the test proton (red and green lines, respectively) and the evolution
of the particle energy (blue line).  In the right panel it is shown the
change of the particle kinetic energy with the X coordinate.  The proton
orbiting around the center of the magnetic island increases its energy increment
after each orbit. \label{fig:event}}
\end{figure*}

In Figure~\ref{fig:event}, we show an example of a test proton which is trapped
in a contracting island and accelerates increasing its parallel component.  The
left panel of the figure shows the topology of the magnetic field in a region
with a contracting island (R1 in Fig.~\ref{fig:locations}) with the trajectory
of the proton superimposed.  As long as the proton remains trapped in the island
it orbits around the island center.  In the middle panel we plot the evolution
of the parallel and perpendicular velocity components (red and green lines,
respectively), and the kinetic energy evolution (blue line) of the test
particle.  In the right panel, we plot the change of the kinetic energy with the
particle X coordinate.  While trapped in the island, the particle orbits around
the center and increases its energy after each cycle.  Its parallel speed
increases while the gyro rotation slows down.  This results in an exponential
growth of the kinetic energy of the particle (see middle panel).  If we take a
closer look in the change of the kinetic energy with the position we see that
the increase of energy happens only when the particle moves across the right
part of the island.  The left panel showing the magnetic field topology
indicates that the field lines in this part are contracting due to the
interaction with a small island which is merging with the central one.  At the
same time, the particle gains more energy after each pass in this zone.

We clearly see how easily the results of \cite{drake10} can be reproduced using
the MHD approximation, confirming that the process of acceleration in the
islands is not restricted to the collisionless physics described by PIC codes.
MHD codes present an easier way to study the physics of particle acceleration
numerically.

\subsection{Acceleration Near and Within Current Sheets}
\label{sec:current_sheets}

We know from shock acceleration theory that particles are injected upstream and
allowed to convect into the shock, while diffusing in space so as to effect
multiple shock crossings, and thereby gain energy through the first order Fermi
process.  Now, besides this mechanism, they may also experience shock drift
acceleration, which is attributed to the grad-B drift when the particle
encounters an abrupt change in magnetic fields.  The origin of this effect is
due to the net work done on a charge by the Lorentz force (Eq.~\ref{eq:efield})
in a zone of non-uniform large scale magnetic field.  The principal equation
governing this is the scalar product of the particle velocity (or momentum) and
the acceleration by the convective electric field, $-q \vc{v} \times  \vc{B}$.
In uniform magnetic fields, the energy gain and loss acquired during a
gyroperiod exactly cancel, so in result no net work is done, $\Delta W = 0$.

In contrast, when the gyromotion of a charged particle straddles a
discontinuity, the sharp field gradient induces an asymmetry in the time spent
by the charge in either side of the discontinuity, so that energy gain and loss
do not compensate each other.  The compressive nature of the field discontinuity
biases the net work done to positive increments in shock encounters between
upstream excursions, and it can be shown that $\Delta W = q E_x dx$, where the
$\vc{v} \times \vc{B}$ drifts lie in the x-direction, in other words, this
energy gain scales linearly with displacement along the drift coordinate x
(i.e., along the discontinuity).  Thus the energy gained by a particle depends
on how far it drifts {\em along} the front \cite[see, e.g.,][and references
therein]{baring09,lugones11}.

\begin{figure*}[ht]
 \center
 \includegraphics[width=0.40\textwidth]{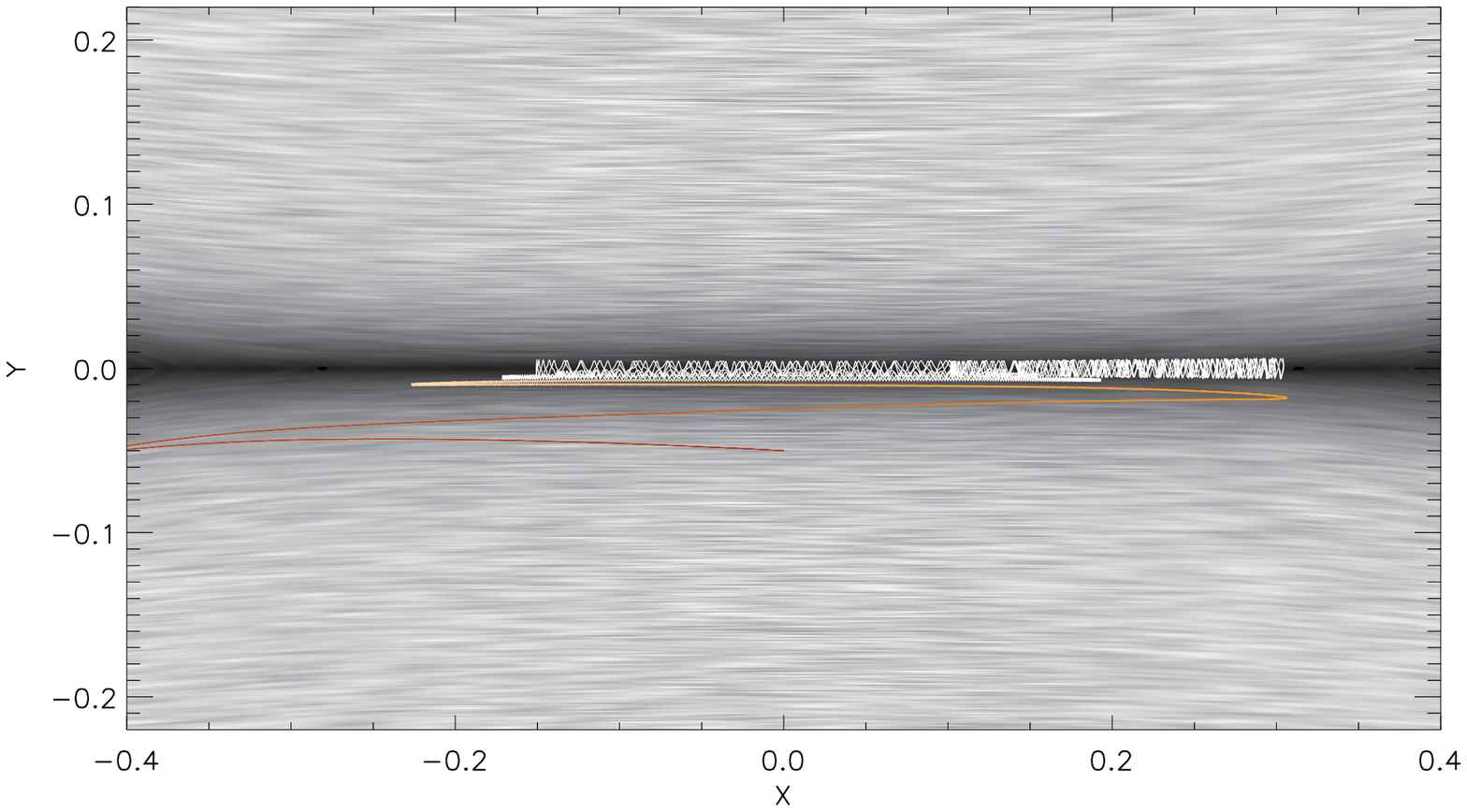}
 \includegraphics[width=0.30\textwidth]{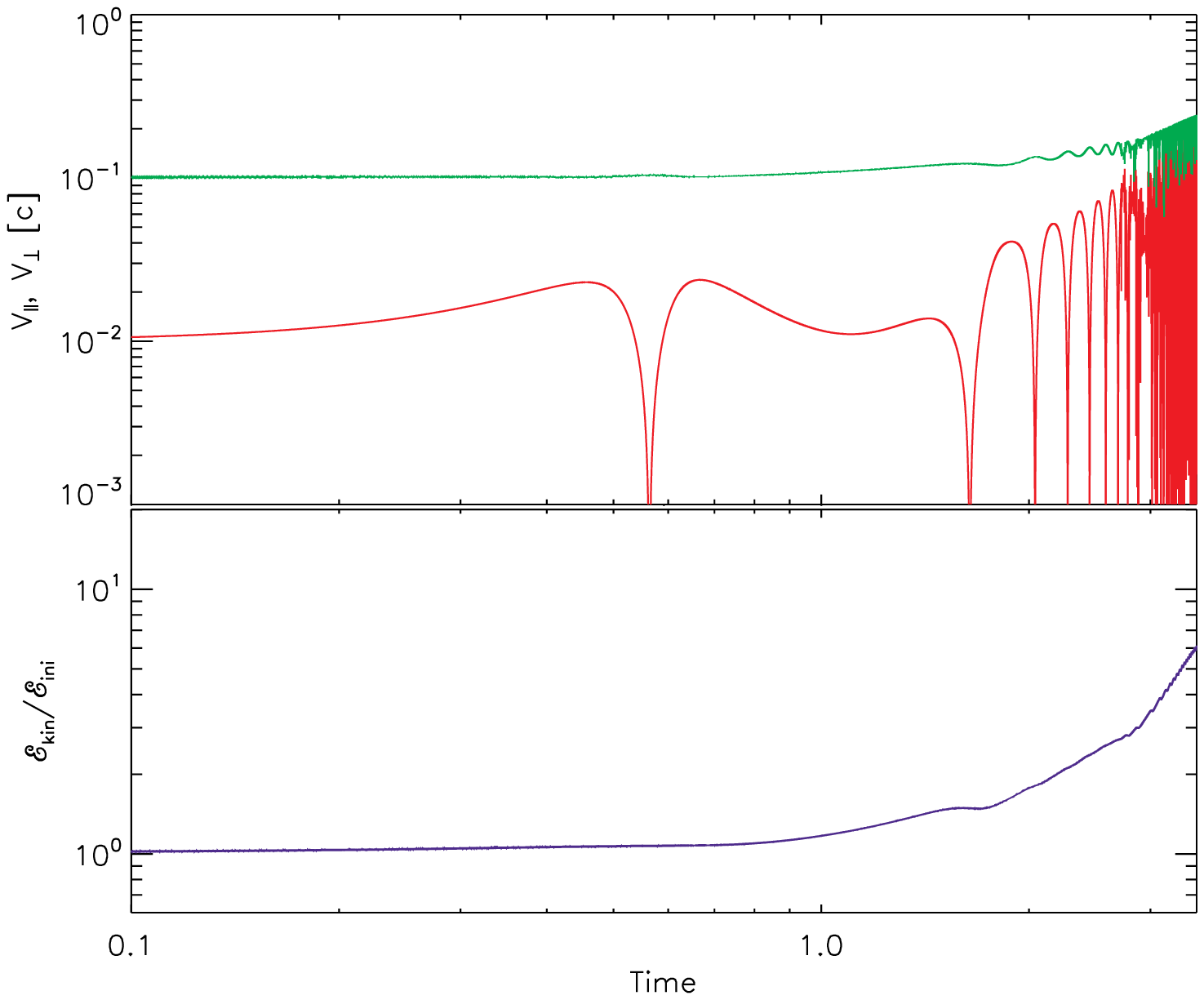}
 \includegraphics[width=0.25\textwidth]{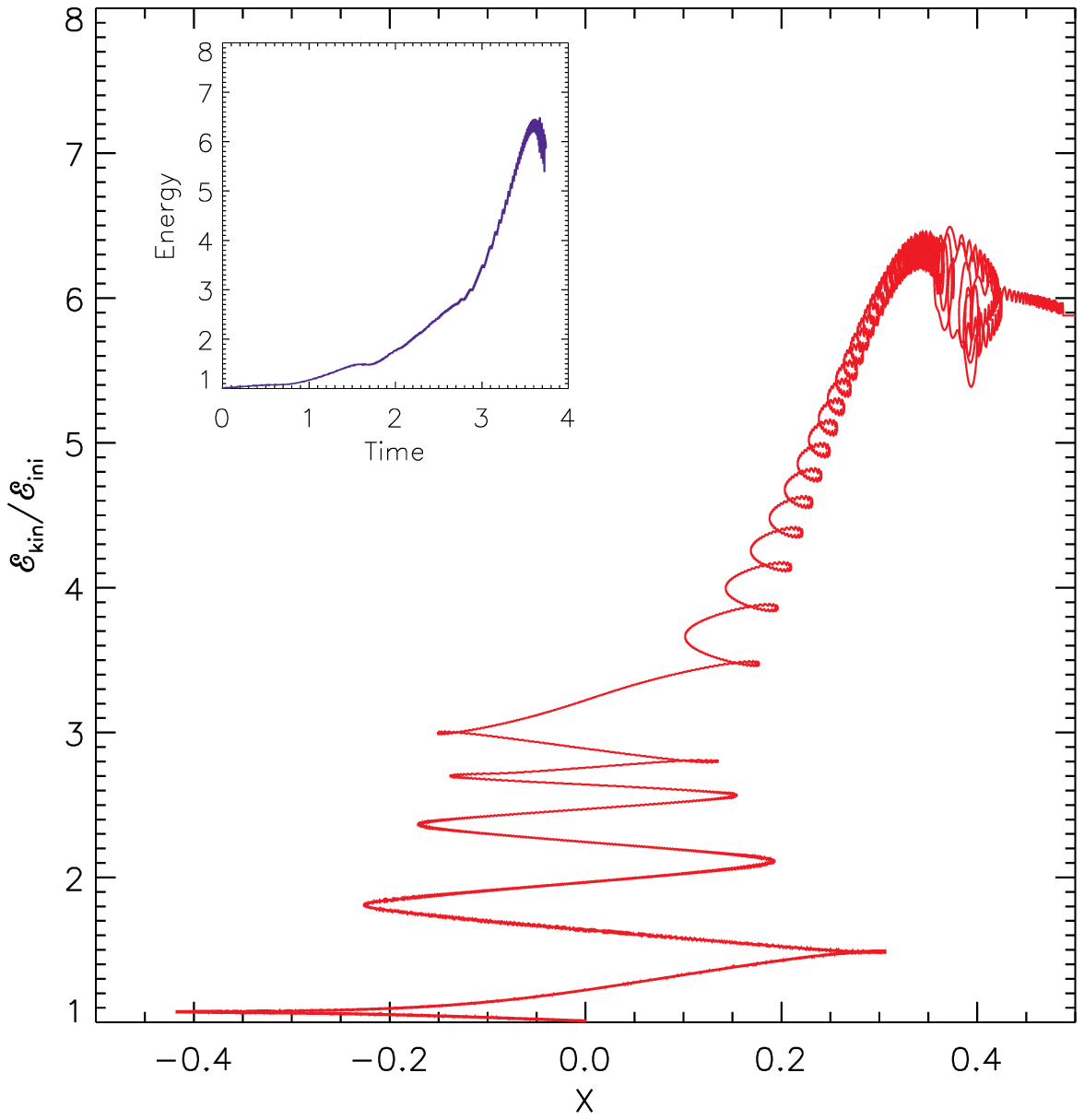}
 \caption{The case of acceleration near and within a single current sheet with a
Sweet-Parker configuration where the particles accelerate efficiently (like
region R6 of Fig.~\ref{fig:locations}).  The left panel shows the trajectory of
a test proton approaching the diffusion region.  The color of trajectory
corresponds to the particle energy (which increases from red to yellow and then
finally to white when the particle reaches the current sheet). The middle panel
shows the evolution of the parallel and perpendicular speeds of the test proton
(red and green lines, respectively) and the evolution of the particle energy
(blue line).  The right panel shows the change of the particle kinetic energy
with the X coordinate.  Outside the current sheet the proton drifts under the
effect of the magnetic field gradients and then when it arrives in the current
sheet it bounces back and forth between the converging magnetic fluxes of
opposite polarity while drifting along the magnetic field.
\label{fig:sp_acceleration}}
\end{figure*}

Similarly, in the current sheet zones (regions R4 to R6) we may be identifying
two distinct mechanisms of acceleration, either first order Fermi acceleration
in contracting/merging islands which are just forming there \cite[or even the
simple particle scattering between the converging flows entering both sides of
the current sheet, as described in][]{degouveia05}, or a drift acceleration, as
described above.  The islands in current sheets are smaller than those of
regions R1, R2 and R3, what results in smaller acceleration rates as we see in
the lower panels of Figure~\ref{fig:particle_histograms}.  In the zones above
and below the current sheets it is possible that we see predominantly a drift
acceleration driven by non-uniformities of the magnetic field.  Generally, this
effect is less efficient than the first order Fermi process in
merging/contracting islands and results in smaller acceleration rates.

In order to better understand these distinct acceleration mechanisms, we have
also explored the details of the acceleration of a test particle near and within
a single (Sweet-Parker shaped) current sheet.  Figure~\ref{fig:sp_acceleration}
shows the trajectory and energy evolution of this test particle.  We see that
before the particle reaches the current sheet discontinuity it is drifted by the
plasma inflow and the increasing gradient of B as it approaches the current
sheet.  When it enters the discontinuity (the white part of the trajectory in
the left panel), it bounces back and forth several times and gains energy (which
increases exponentially as shown in the middle panel of
Figure~\ref{fig:sp_acceleration}) due to head-on collisions with the converging
flow, on both sides of the magnetic discontinuity \cite[i.e., in a first order
Fermi process, as described in][]{degouveia05}.  At the same time it drifts
along the magnetic lines which eventually allow it to escape from the
acceleration region.  Therefore, we see two mechanisms: a drift acceleration
(dominating outside of the current sheet) and first order Fermi acceleration
inside the current sheet.  These processes naturally depend on the initial
particle gyroradius, since it determines the amount of time the particle remains
in the acceleration zone before escaping.

Finally, we may also argue that turbulence above and below the current sheets in
Figure~\ref{fig:locations}, far from the islands and diffusion regions, favor
second order Fermi acceleration mechanisms with particles being scattered by
approaching and receding magnetic irregularities.  Nevertheless, the first order
Fermi processes occurring within the islands and current sheets dominate the
overall particle acceleration in the system.

\subsection{The Role of a Guide Field: 2D vs. 3D simulations}
\label{sec:2d_vs_3d}

The results presented in the previous sections were obtained for 2D models
without a guide field.  This means that in this case the magnetic lines creating
the islands are closed and a charged particle can be trapped indefinitely in
such an island.  The presence of a guide field normal to the plane of
Figure~\ref{fig:locations} opens the magnetic loops and allows the charged
particles to travel freely in the out-of-plane direction.  Moreover, the islands
evolve much slower in the presence of a strong guide field.

\begin{figure}[ht]
 \center
 \includegraphics[width=0.5\textwidth]{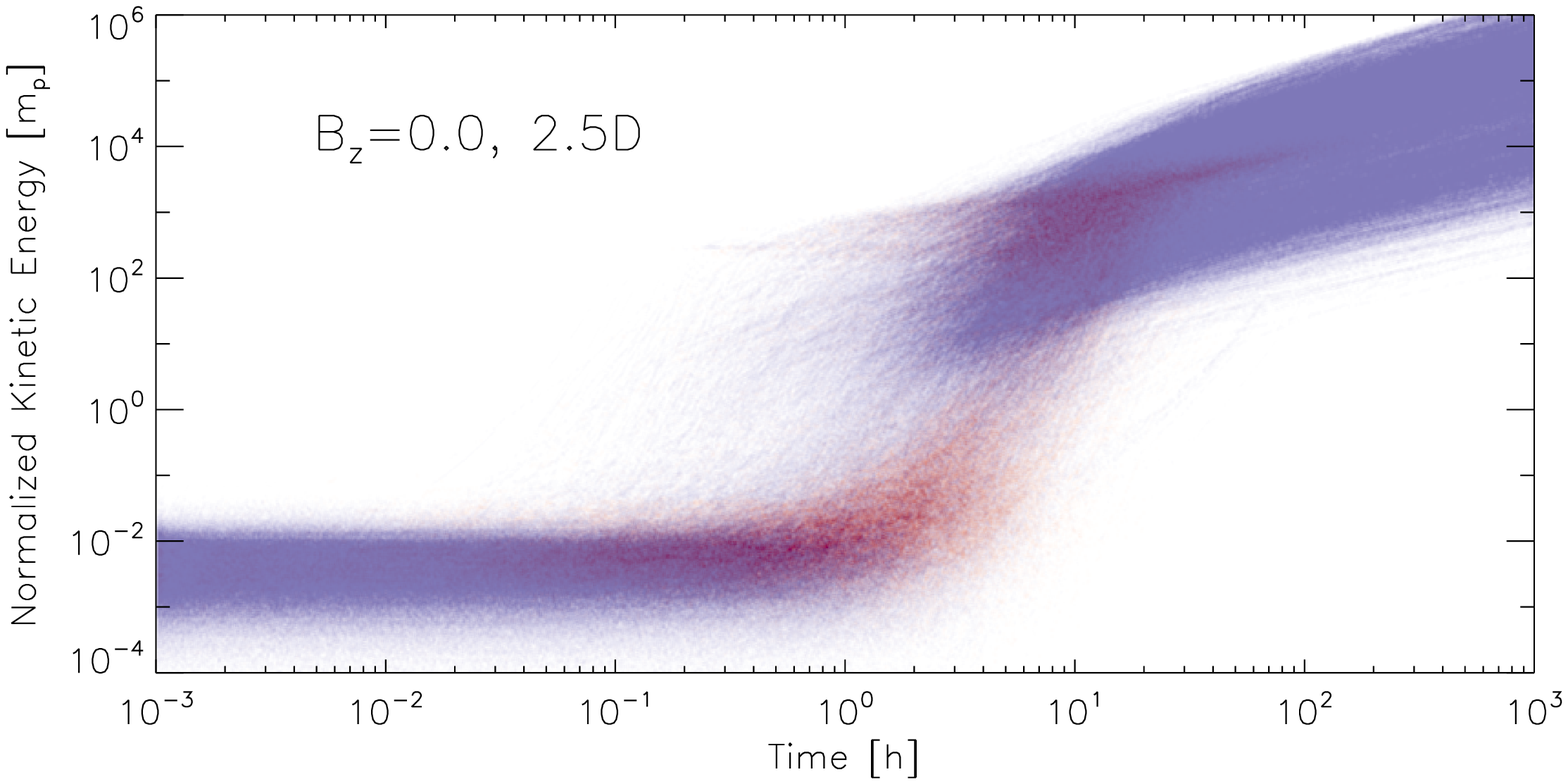}
 \includegraphics[width=0.5\textwidth]{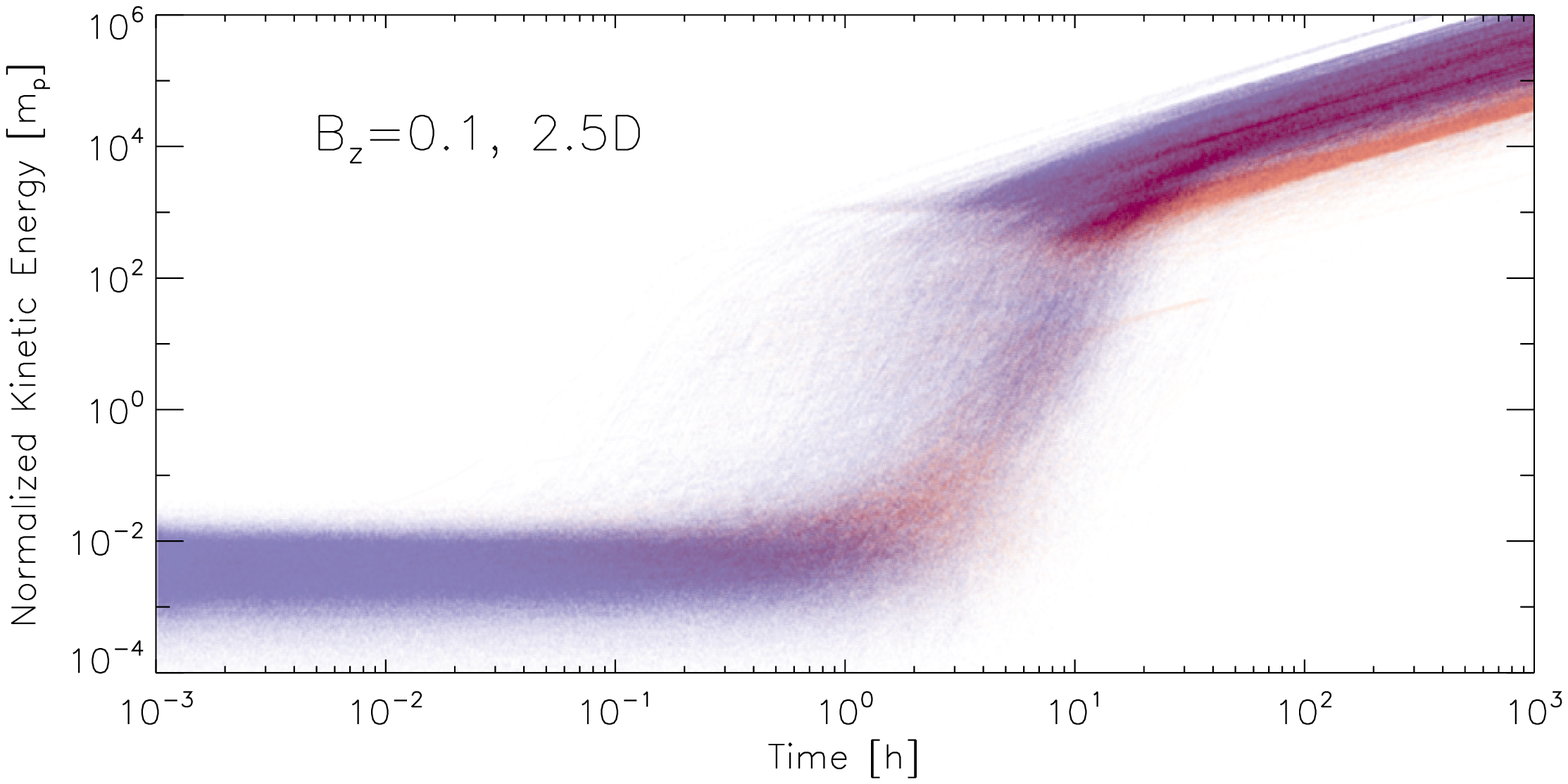}
 \includegraphics[width=0.5\textwidth]{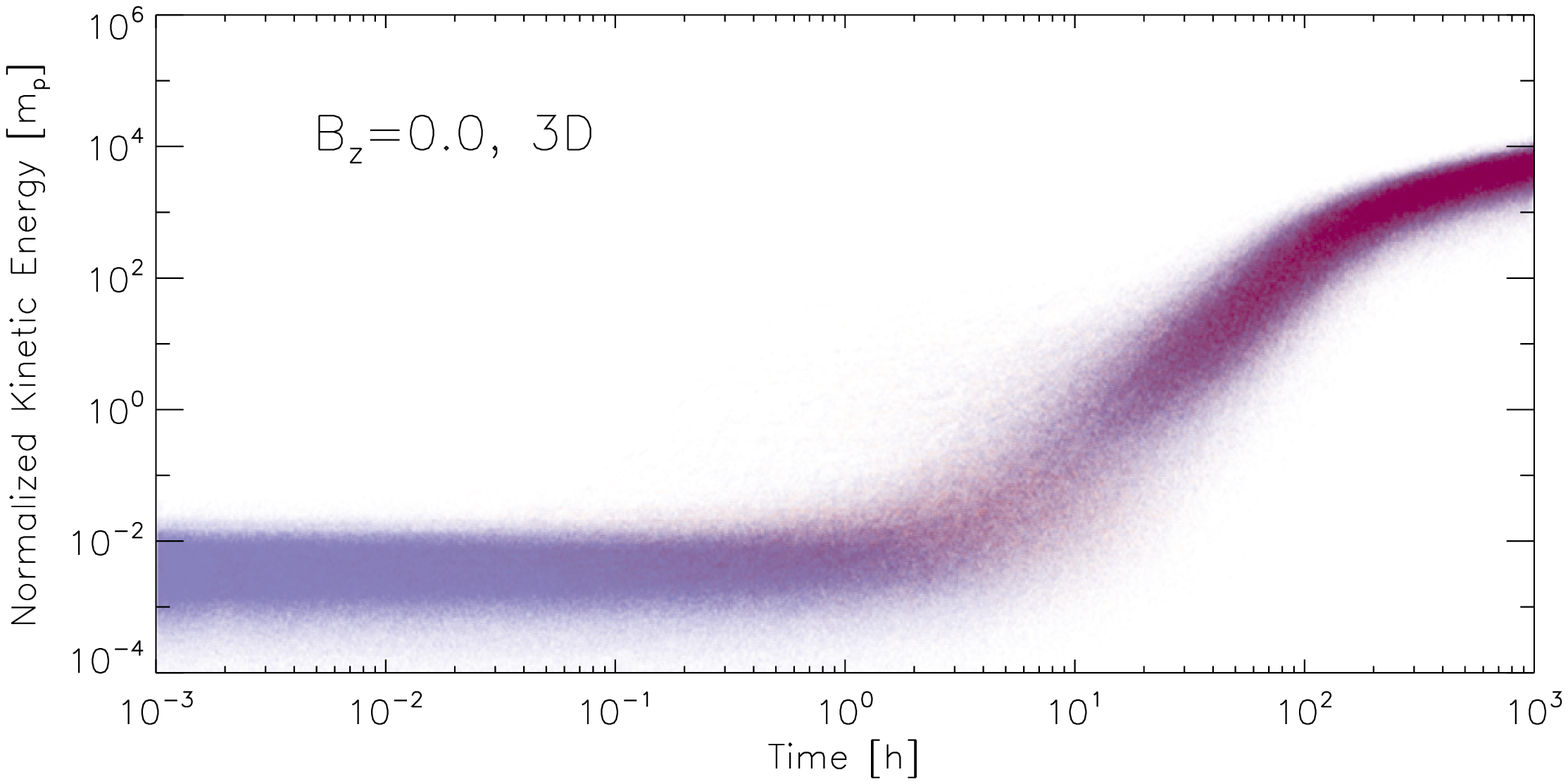}
 \caption{Kinetic energy evolution of a group of 10$^4$ protons in 2D models of
reconnection with a guide field strength $B_z$=0.0 and 0.1 (top and middle
panels, respectively).  In the bottom panel a fully 3D model with initial
$B_z$=0.0 is presented.  The colors show how the parallel (red) and
perpendicular (blue) components of the particle velocities increase with time.
The energy is normalized by the rest proton mass energy.  The background
magnetized flow with multiple current sheet layers is at time 4.0 in Alfv\'en
time units in all models.  \label{fig:energy_2d_3d}}
\end{figure}

In Figure~\ref{fig:energy_2d_3d}, we present the time evolution of the kinetic
energy of the particles which have their parallel and perpendicular (red and
blue points, respectively) velocity components accelerated for three models of
reconnection. The upper panel shows the energy evolution for a 2D model without
the guide field (as in the models studied in the previous sections).  Initially,
the particles pre-accelerate by increasing their perpendicular velocity
component only.  However, later there is an exponential growth of energy mostly
due to the acceleration of the parallel component which stops after the energy
reaches values of 10$^3$--10$^4$~$m_p$ (where $m_p$ is the proton rest mass
energy).  From that level on, particles accelerate their perpendicular component
only with smaller linear rate in a log-log diagram.  The middle panel shows the
kinetic energy evolution in a 2D model with a weak guide field $B_z$=0.1 normal
to the plane of Figure~\ref{fig:locations}.  In this case, there is also an
initial slow acceleration of the perpendicular component followed by the
exponential acceleration of the parallel velocity component. However, due to the
presence of a weak guide field, the parallel component accelerates further to
higher energies at a similar rate as the perpendicular one. This implies that
the presence of a guide field removes the restriction seen in the 2D model
without a guide field and allows the particles to increase their parallel
velocity components as they travel along the guide field, in open loops rather
than in confined 2D islands. This result is reassured by the 3D model in the
bottom panel of Figure~\ref{fig:energy_2d_3d}, where no guide field is necessary
as the MHD domain in fully three-dimensional. In this case, we clearly see a
continuous increase of both components, which suggests that the particle
acceleration behavior changes significantly when 3D effects are considered,
where open loops replace the closed 2D reconnecting islands.

Considering the parametrization we have chosen for our models, the gyroradius of
a proton becomes comparable to the size of the box when its Lorentz factor
reaches a value of a few times 10$^4$.  The largest islands in the system can
have sizes of a few tenths of the size of the box.  These rough estimates help
us to understand the energy evolution in Figure~\ref{fig:energy_2d_3d} and the
transition from an exponential to a much slower (linear) growth rate in the
energy around 10 hours.  We note that in the diagrams of this figure, the energy
is normalized by the rest mass value, so that in fact, it is the Lorentz factor
that is plotted.  In the case with absence of a guide field (top panel of
Fig.~\ref{fig:energy_2d_3d}), the exponential parallel acceleration stops right
before the energy value 10$^4$ is reached.  After this, the rate of acceleration
significantly decreases.  This occurs because the Larmor radius of the particles
has become larger than the sizes of biggest islands.  Therefore, from this level
on the particles cannot be confined anymore within the islands and the first
order Fermi acceleration ceases.  After that, there is a much slower drift
acceleration (of the perpendicular component only) caused by the gradients of
the large scale magnetic fields.  If a guide field is inserted in such a system,
the picture is very similar, except for one detail.  Now, since the particles
are able to travel along the guide field, their parallel velocity component also
continues to increase after the 10$^4$ threshold (see the middle panel of
Fig.~\ref{fig:energy_2d_3d}).  Of course, in the 3D model, the particles follow
the same trend (bottom panel of Fig.~\ref{fig:energy_2d_3d}).

\begin{figure}[ht]
 \center
 \includegraphics[width=0.5\textwidth]{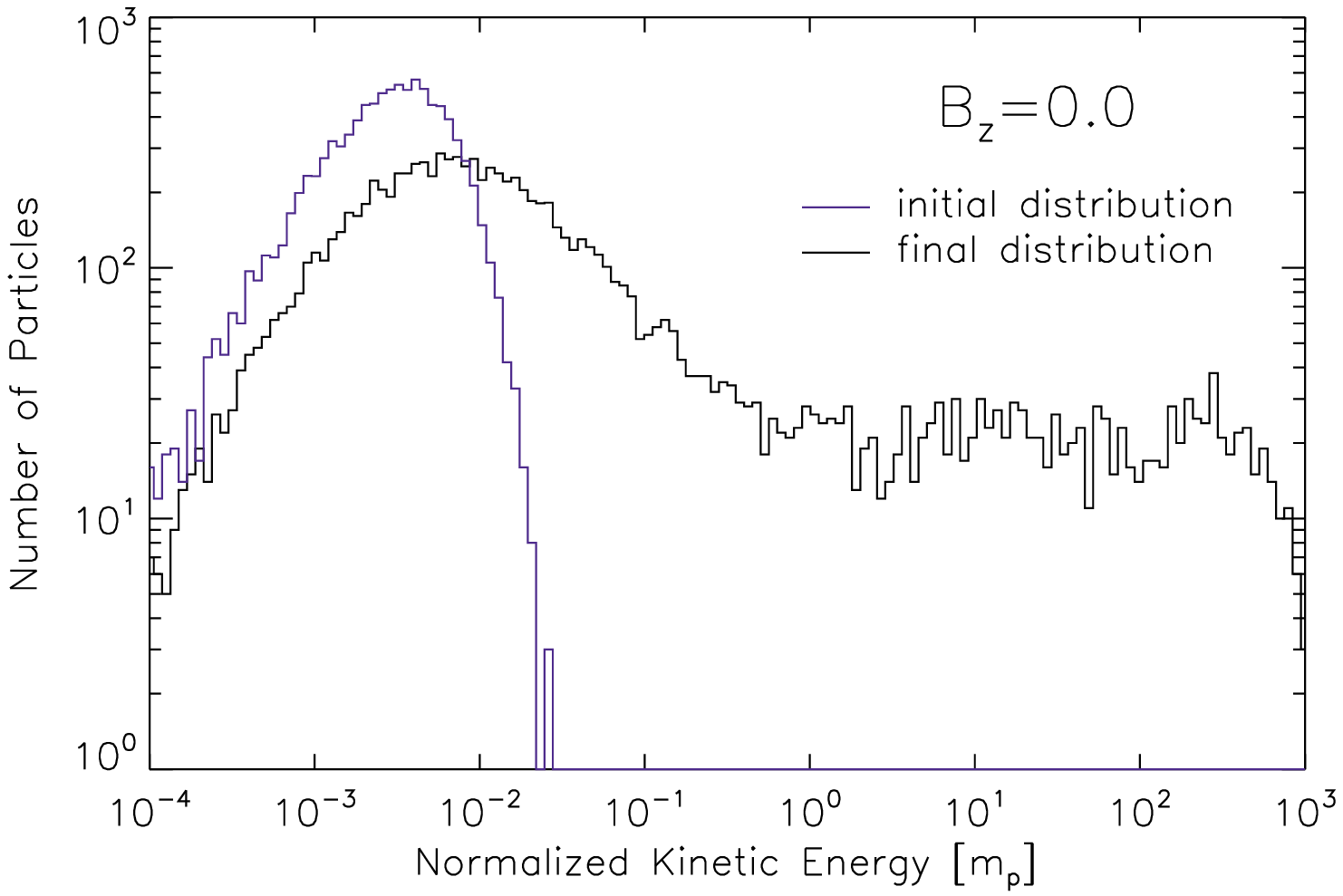}
 \includegraphics[width=0.5\textwidth]{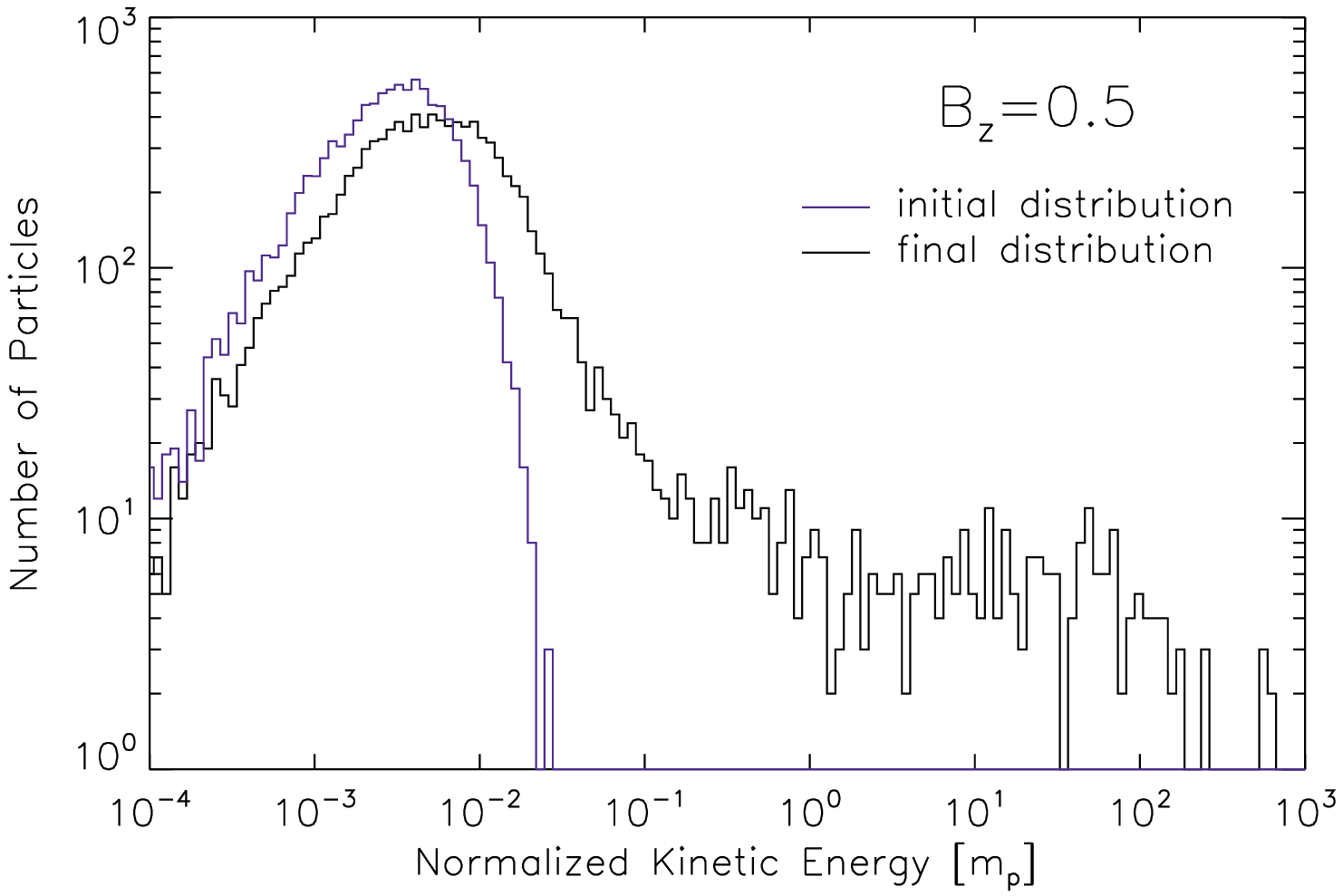}
 \includegraphics[width=0.5\textwidth]{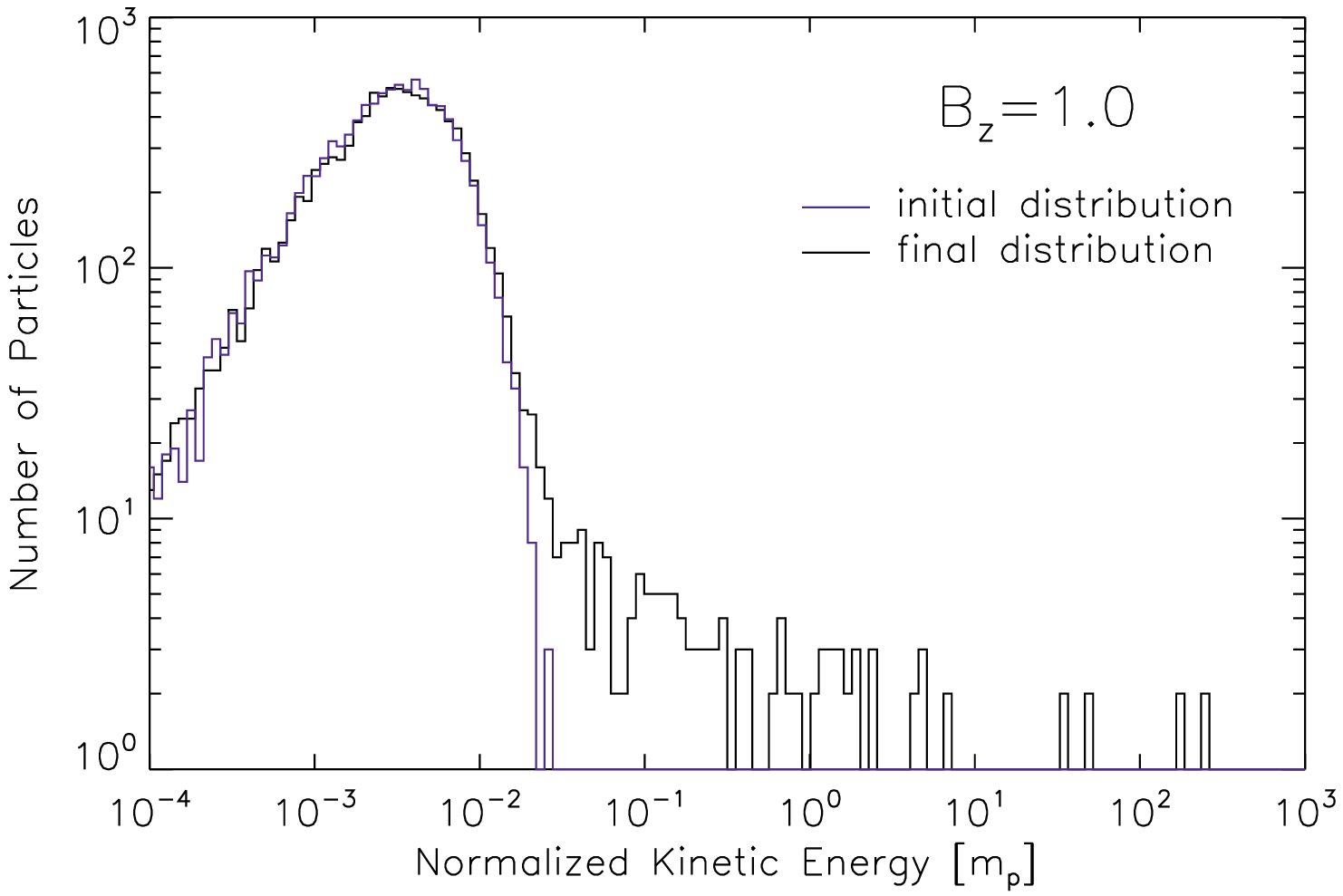}
 \caption{Kinetic energy distribution of the particles for three 2D models with
different strengths of the out-of-plane guide field $B_z$ = 0.0, 0.5, and 1.0
(top, middle and bottom panels, respectively.) at a time corresponding to 1 hour
after the injection.  The initial particle thermal distribution is given by the
blue line. \label{fig:energy_guide_field}}
\end{figure}

Figure~\ref{fig:energy_guide_field} exhibits the dependence of the acceleration
rate on the out-of-plane guide field strength. It shows the particle kinetic
energy distribution for three models of reconnection with different strengths of
the guide field ($B_z$=0.0, 0.5, and 1.0 for the top, middle, and bottom panels,
respectively) 1 hour after the injection.  The initial injected thermal
distribution is represented by the blue line.  The particles trajectory
integration was performed at the same MHD snapshot, at time 4.0 (in units of
Alfv\'en time), for all particles in these models.  We note that at the final
state the particle distributions have developed high energy tails which depart
from the initial thermal distribution. However, both the number of accelerated
particles  to higher energies and the maximum energy at the final time interval
strongly depends on the strength of the guide field.

\section{Discussion}
\label{sec:discussion}

In this paper we have investigated the process of particle acceleration in
reconnection zones considering test particles inserted in an isothermal
magnetohydrodynamic domain containing a series of reconnection layers, without
including kinetic effects.  There were a few goals that motivated our approach.
The first goal was to study if the contraction of magnetic islands, which
develop in 2D reconnection layers with finite resistivity, can also occur when
the gas pressure is isotropic, as is the case in the MHD regime.  In earlier
studies, Drake and collaborators \citep{drake10} invoked kinetic effects such as
the firehose and mirror instabilities to control the development and contraction
of the islands in current sheets.  In the present work, after obtaining the
formation of contracting/deformed islands in current sheets in a nearly
incompressible MHD domain, the second goal was to investigate the acceleration
of test particles particularly in the direction parallel to the large scale
magnetic fields during island contractions.  We were able to identify several
regions where such acceleration occurs both in a 2D environment with and without
a guide field and in a 3D environment.  Also, we were able to identify the
nature of the acceleration mechanisms in the different acceleration regions, not
only within the contracting islands, but also near and inside the current
sheets.  We found that first order Fermi processes occur in the islands and also
within the current sheets, while drift acceleration due to magnetic field
gradients seems to be dominating outside the current sheets.

Our results for particle acceleration in 2D MHD models were quite similar to
those obtained by \cite{drake10}, i.e., during an island contraction, in our
case resulting from the merge with other islands, a particle trapped in it can
accelerate and increase its energy exponentially and its parallel velocity
component grows while the perpendicular one undergoes a net decrease.  During
this process the energy gain is proportional to the particle energy.  These are
all necessary conditions in a Fermi like process and a close examination of the
motion of a test particle in a contracting island shows clear evidence of the
first order Fermi acceleration process.

Similarly, exponential acceleration of the parallel velocity component was also
detected in regions within the current sheets.  A detailed examination of a test
particle trajectory within and near a current sheet revealed that outside the
discontinuity the particle experiences drift acceleration due to magnetic field
gradients and within the current sheet it goes back and forth between the
converging flows on both sides of the discontinuity also gaining energy
exponentially, just as described in \cite{degouveia05}, while drifting along the
magnetic field direction.

\cite{drake10} have claimed that particle acceleration in an X-point
reconnection in a 3D environment, even in the presence of a guide field, should
behave like in 2D systems.  However, we have shown that in the presence of an
out-of-plane guide field, this picture changes completely \cite[see
also][]{drake10}.  While in 2D MHD models without a guide field the parallel
acceleration stops after reaching a certain energy threshold, in 2D MHD models
with a guide field this constraint is removed, and particles can continue
increasing their parallel speed as they travel along the guide field.
Furthermore, in fully 3D MHD models with no guide field the acceleration of the
particles exhibits the same trend as in 2D models with a guide field, i.e.,
there is no constraint on the acceleration of the parallel speed.  This result
is very important as it implies that the overall picture of particle
acceleration in 3D reconnection can be very distinct from that in 2D
reconnection.  Also, it can offer some important clues for particle acceleration
in more complex domains, such as in the presence of 3D turbulence where
stochastic reconnection is involved \cite[see][in prep.]{kowal11}.

Another important result of the present work was to demonstrate that the
investigated acceleration mechanisms in reconnection sites (both drift and Fermi
processes) can be present in any environment whose evolution can be approximated
by a nearly ideal MHD description as employed here, therefore, without the
necessity of invoking kinetic instabilities or anomalous resistivity effects to
control the pressure anisotropy or the reconnection rate.  The nearly
non-resistive MHD approach with test particle injection offers a possibility of
exploring more realistic systems.  Even though PIC codes in general allow the
study of plasma processes in greater detail, they present some disadvantages
with regard to an MHD description, such as the necessity of a much higher
complexity in the 3D modeling, for instance, of turbulence.

A final remark is in order.  \cite{onofri06} investigated particle acceleration
in reconnection zones and concluded that MHD should not be a good approximation
to describe the whole process of acceleration.  However, their 3D numerical
simulations were performed in a fully resistive MHD regime. Therefore,  they
obtained very efficient particle acceleration due to the high electric field
induced by resistivity and an absorption of most of the available magnetic
energy by the electrons in a very small fraction of the characteristic time of
the MHD simulation.  This led them to conclude that resistive MHD codes are
unable to represent the full extent of particle acceleration in 3D reconnection.
 In the present work, we have investigated particle acceleration in a nearly
ideal MHD regime where only numerical resistivity is present.  In this case, the
contribution of a resistivity induced electric field is negligible when compared
to the advection component, namely, the electric field resulting from the plasma
motion in the magnetized medium, $\vc{E} = - q \vc{v} \times \vc{B}$.  Moreover,
since there is no important dissipation effects, our model is scale independent
and allows for dimensionless units, so that it is straightforward to rescale the
configuration to reproduce different astrophysical environments.  The main
drawback of the numerical resistivity is the creation of a ''hole'' in the
center of the magnetic islands, since the numerical modeling cannot handle
properly the strongly curved magnetic lines over a few cells.  As a consequence,
numerical dissipation removes part of the magnetic flux in this region.

The importance of the present study of acceleration in the MHD regime is
motivated by the fact that magnetic reconnection in 3D becomes fast according to
the model in \cite{lazarian99} \cite[see also][]{lazarian04}.  This model has
been successfully tested via numerical simulations in \cite{kowal09} which
confirmed that thick reconnection layers form, where magnetic energy is
transferred into energy of contracting loops \footnote{These thick layers were
also confirmed by \cite{ciaravella08}.}.  In the present paper, we have
confirmed that these loops (or magnetic irregularities) can act as the places of
efficient particle acceleration providing the support for the mechanism first
discussed in \cite{degouveia05}.  At the same time, we see that the process of
acceleration does not amount to only the process of first order Fermi
acceleration outlined in the aforementioned paper.  Our simulations show a
complex interplay of different acceleration processes, which motivates for
further studies of the acceleration in the presence of realistic turbulent
reconnection \cite[see][]{kowal09}.

The present study has clearly an exploratory character.  It testifies that the
acceleration in reconnection regions may be more complex than it may be inferred
from earlier simplified 2D studies.  Further study of the acceleration in the
MHD regime is absolutely essential, as ubiquitous astrophysical turbulence is
expected to induce magnetic reconnection all over astrophysical volumes and
should be explored in more detail.  In this situation the acceleration of
particles by reconnection may play a vital role, the extend of importance of
which can be evaluated from further research \citep{kowal11}.

\section{Summary}
\label{sec:summary}

The results of the paper can be very briefly summarized as follows:
\begin{itemize}
 \item Advances in the understanding of magnetic reconnection in the MHD regime,
in particular, turbulent magnetic reconnection in \cite{lazarian99} model
motivate the studies of whether the reconnection in this regime can accelerate
energetic particles.

 \item Contracting magnetic loops in magnetic reconnection in 2D, in the MHD
regime, provides the acceleration which successfully reproduces the results
obtained earlier with more complicated PIC codes, which proves that the
acceleration in reconnection regions is a universal process which is not
determined by the details of plasma physics.

 \item Acceleration of energetic particles in 2D and 3D shows substantial
differences, which call for focusing on realistic 3D geometries of reconnection.
 Our study also shows that apart from the first order Fermi acceleration,
additional acceleration processes interfere.
\end{itemize}

\acknowledgements

GK and EMGDP acknowledge the support by the FAPESP grants no. 2006/50654-3 and
2009/50053-8, and the CNPq  grant no. 300083/94-7, and AL thanks the NSF grant
AST 0808118, NASA grant NNX09AH78G and the support of the Center for Magnetic
Self Organization. This research was also supported by the project TG-AST080005N
through TeraGrid resources provided by Texas Advanced Computing Center
(TACC:http://www.tacc.utexas.edu).  Part of the computations presented here were
performed on the GALERA supercomputer in the Academic Computer Centre in
Gda\'nsk (TASK:http://www.task.gda.pl/).


\end{document}